\documentclass[aps,prd,twocolumn,groupedaddress,showkeys]{revtex4}
\usepackage{mathrsfs}
\usepackage{amssymb}
\usepackage[intlimits]{amsmath}
\usepackage{relsize}
\usepackage[edtable]{lineno}
\usepackage{comment}
\usepackage{epsfig}
\begin{document}

% Use the \preprint command to place your local institutional report
% number in the upper righthand corner of the title page in preprint mode.
% Multiple \preprint commands are allowed.
% Use the 'preprintnumbers' class option to override journal defaults
% to display numbers if necessary
%\preprint{}

%Title of paper
\title{Quantum Field Theory and the Internal States of Elementary Particles}
%\title{Quark Modes as Solutions of the QCD Equations }
% repeat the \author .. \affiliation  etc. as needed
% \email, \thanks, \homepage, \altaffiliation all apply to the current
% author. Explanatory text should go in the []'s, actual e-mail
% address or url should go in the {}'s for \email and \homepage.
% Please use the appropriate macro foreach each type of information

% \affiliation command applies to all authors since the last
% \affiliation command. The \affiliation command should follow the
% other information
% \affiliation can be followed by \email, \homepage, \thanks as well.

\author{J. M. Greben}
\address{CSIR, PO Box 395, Pretoria 0001, South Africa }

%\email[]{Your e-mail address}
%\homepage[]{Your web page}
%\thanks{}
%\altaffiliation{}

%Collaboration name if desired (requires use of superscriptaddress
%option in \documentclass). \noaffiliation is required (may also be
%used with the \author command).
%\collaboration can be followed by \email, \homepage, \thanks as well.
%\collaboration{}
%\noaffiliation

%\date{\today}

\begin{abstract}
A new application of quantum field theory is developed that gives
a description of the internal dynamics of dressed elementary
particles and predicts their masses. The fermionic
and bosonic quantum fields are treated as interdependent fields
satisfying coupled quantum field equations, all
expressed at the same space-time coordinate. Quantization is
realized by expanding the quantum fields in terms of fermionic creation
and annihilation operators. This approach is applied in a QCD description of
the light quarks with a zero Higgs field. Originally massless and pointlike,
an isolated quark (described in its own center-of-mass) acquires mass and a finite extent when treated as an
interacting system of quark and gluon fields. The binding mechanism of this localized system has a topological character, being a consequence of the non-linear nature of QCD, while being insensitive to the magnitude of
the coupling constant to lowest order. To prevent this system from collapsing general relativity is introduced. The quark stabilizes at a radius of 8.8 Planck lengths and acquires a mass of 3.2 MeV, in remarkable agreement with accepted phenomenological values. It is suggested that the two higher generations of quarks are associated with the other two real solutions of the Higgs field equations.

\end{abstract}

% insert suggested PACS numbers in braces on next line
%\pacs{12.38.Lg, 11.10.Lm, 11.15.Tk, 14.65.Bt}
% \pacs{11.10-q, 11.15.Tk, 12.38-t and 12.38.Lg} as submitted but corrected in application
% insert suggested keywords - APS authors don't need to do this
%\maketitle must follow title, authors, abstract, \pacs, and \keywords
\maketitle

% body of paper here - Use proper section commands
% References should be done using the \cite, \ref, and \label commands
\section{Introduction}

\label{sec:Introduction}
% Put \label in argument of \section for cross-referencing
%\section{\label{}}
The standard model has been extremely successful in describing scattering phenomena
between elementary particles. Nonetheless, it is clear that this picture of Nature cannot
be complete. Most importantly, the standard model contains too many
free parameters to be called fundamental, and can thus best be seen as an effective theory.
In modern treatments the masses of the standard model fermions are supposed to arise from the couplings of quarks to various Higgs fields, however, so far this has not lead to a reduction of the number of parameters. Of the 19 free parameters of the standard model, 13 belong to the Yukawa/Higgs sector \cite {Gabrielli}. These parameters are put in by hand, and there is no explanation for the huge hierarchy of masses ranging over 6 orders of magnitude (not including the neutrinos). Further extensions beyond the standard model often lead to a greater - rather than a smaller - set of unknown parameters.

Given this situation one may wonder whether it is not possible to underlie the standard model by a theory at a more fundamental level that evolves into the standard model through the construction of effective solutions of the bare degrees of freedom. Such an underlying theory should contain many of the elements of the standard model but have a simpler structure, i.e. it should embody a certain unification of the diverse elements of the effective theory. The simplicity should imply fewer degrees of freedom and fewer parameters at the fundamental level. The new application of QFT methods, introduced in this paper, enables such a link between fundamental and effective levels by its ability to construct massive dressed elementary particles from more fundamental bare massless particles.

An important element of such a fundamental theory could be the reduction of the three generations in the effective theory to a single one in the fundamental theory. Since the three generations in the standard model are not associated with any additional symmetry or interaction, three generations may well be the result of multiple solutions at a more fundamental level.  This possible scenario would gain in standing if we could explain the emergence of (exactly) three generations from the exact level. We suggest that a single fundamental Higgs field might explain the threefold split. The Higgs field equations would typically have three real solutions, whose classical approximations could be denoted by $\phi=0$ and $\phi =\pm \sqrt{|\mu^2/\lambda|}$. In the quark-Higgs dynamics the positive and negative solutions lead to distinct solutions, so that these three solutions would indeed lead to three distinct fermion sectors. Since the nature of the Higgs Lagrangian at the fundamental level is not (yet) unambiguously established and a non-zero Higgs field would complicate our basic model considerably, we limit ourselves in this paper to the trivial Higgs solution, namely $\phi=0$. However, this also eliminates the Higgs parameter $\mu$ from the model, so that it is unclear how this basic theory acquires a scale. The question of the basic scale parameters in Nature has been considered in the context of cosmology elsewhere \cite {Greben}.  We find that for the light quarks general relativity has to be introduced to ensure the existence of the quarks, so that for this basic generation the cosmological parameters reviewed in \cite {Greben} also play a role. The presence of different scale parameters at this fundamental level could give a possible explanation of the hierarchy problem in physics, as it would allow for the existence of a range of different scales at a very basic level.

Which further ingredients would survive in a fundamental theory underlying the standard model? It is to be expected that the SU(3) theory of strong interactions will survive, as this theory is unbroken at the standard model level and is not characterized by a complex multiplet structure. The complexities of the multiplet structure and broken symmetries in the standard model originate from the electro-weak interaction and it is not clear which of these complexities emerge in the transition from fundamental to effective theory, or which ones are already present at the most fundamental level. Therefore it is natural to limit oneself originally to QCD. Clearly, so long as one does not consider the electro-weak interactions one cannot describe leptons as effective systems, so we also focus on quarks in this paper. Being a non-linear theory, one expects that QCD can supply the necessary self-consistent binding mechanism for the effective quark system, recalling the important role non-linear dynamics plays in soliton particle models in hadron physics \cite {Rajaraman}. By adding the electromagnetic interaction at a later stage one can break the isospin symmetry of QCD and try to explain the mass differences between up and down quarks. Being a linear theory, the electromagnetic force can be treated perturbatively and is not expected to interfere with the expected non-linear binding mechanism, so that it can be ignored initially.

As stated before, the possibility to link the dressed particles and their internal properties and masses to more fundamental bare particles relies on a new methodology which emphasizes the use of the field equations in QFT. The standard applications of quantum field theory (QFT) deal with scattering processes expressed in term of propagators and vertices, yielding the well-known Feynman diagrams. The dressing of fermions is then expressed in a series of time-ordered interaction diagrams modifying the bare propagator. Most of these diagrams are divergent, which has necessitated the renormalization methods to arrive at meaningful physical results, although outside inputs are still required to fix the physical parameters. This dressing process does not yield any internal properties of the particles, due to the fact that the fermions interact with external bosonic fields. The dressing process which is introduced in this paper is of a very different nature as the fermionic and bosonic fields are treated as mutually dependent fields, the dependence being controlled by the coupled field equations. Hence, the boson fields are not treated as external fields and all fields and interactions refer to the same space-time coordinate. It is this treatment that enables one to describe the internal properties of elementary particles.

It is clear that the field equations play a dominant role in the new formulation. This is not unlike most other areas of physics. However, in QFT their use has been very limited to date. Often the field equations in QFT are denoted as classical, and indeed they are often used as such. For example, the scalar Higgs field $\phi$ satisfies a classical field equation with various (constant) classical solutions. The "full" solution is then expanded around this classical solution, and only the additional fields are quantized. Despite their classical nature the constant solutions are attributed physical meaning, and their value is often denoted as the vacuum expectation value of the Higgs field \cite {Gabrielli}. This is a dubious use of the term expectation value as it suggests that one takes the expectation value of an operator, which one does not. It is not sufficient to call something an operator, one also has to treat it as such. The expansion around classical solutions is also common in the soliton and instanton problem \cite {Rajaraman}. Here the basic solution also satisfies an approximate non-linear classical equation of motion, and quantization is imposed by linearizing the remainder. We feel that these procedures are fundamentally flawed. If classical solutions have any role to play in QFT then they should emerge as an approximation to the quantum operators, not lie at the basis of the quantization procedure. Quantization should be done up front, not as an afterthought. Hence, we promote a much more rigorous quantum-mechanical use of the field equations. Before solving the field equations one should turn them into quantum equations and ensure that the solutions are quantum field operators.

A rigorous quantization procedure can be realized by turning the classical fields into quantum fields by expanding them into creation and annihilation operators, the latter satisfying well-defined (anti-) commutation rules. The insertion of quantum operators in the field equations also means that the order of the fields in the field equations can make an important difference, a phenomenon well-known in most quantization procedures. Hence, before solving the field equations one must establish the correct order of the fields. This is usually possible by demanding consistency, by imposing symmetries or through other physical requirements. At this point it is of interest to review the standard quantization procedure for a linear field theory such as QED, since this may elucidate why classical solutions have played such a prominent role in traditional quantization procedures, a role which should be avoided when one deals with non-linear equations. To derive the electromagnetic field operator in QED one first determines the classical solutions of the classical linear equation (plane waves in the free case). One then quantizes the field by appending creation and annihilation operators to the classical solutions. This well-known procedure leads to the correct quantized electromagnetic field, and therefore may well be responsible for the practice to use classical solutions in the quantization procedure. However, we would have reached the same result by inverting the order, namely, by first expanding the field in terms of creation and annihilation operators of (as yet) unknown functions, and then solve
the (quantum) equations of motion. This procedure would yield the same plane wave states. However, by choosing this latter order one guarantees up front that the solutions are quantum solutions. For non-linear field equations it makes a huge difference which order is applied. If one solves the non-linear equations classically, and then appends creation and annihilation operators to these solutions, then the resulting operator is generally no longer a solution of the quantum field equations, as the operators do not behave like c-numbers. Hence, one must first introduce the field operators and then construct operator solutions to the field equations. If the resulting profile functions, which multiply the creation or annihilation operators, are identical - or close - to the classical solutions, then this could justify a classical approximation. However, such a situation only arises if the resulting field operator acts like a unit operator in operator space. In general this situation does not apply and classical solutions cannot be expected to play a fundamental role in non-linear field theories.

As indicated above, creation and annihilation operators play a central role in the quantization procedure. The operator nature of the field now even becomes part of the solution process and in general the simple linear expansion of the fermion and boson fields in terms of creation and annihilation operators is no longer a solution of the full set of field equations. One can start with the simple linear expansion of the fermion field, however, after further iterations a much more complex structure will evolve. For example, to lowest order the gluon operator becomes a bilinear operator in terms of the fermionic creation and annihilation operators. This bilinear expansion of the gluon field in terms of fermionic operators, instead of in single bosonic creation and annihilation operators, does not imply that gluon fields have lost their identity. In fact a similar expansion of the gluon field (or the photon field for that matter) in terms of quark operators is possible when it is considered an external field (and thus fully independent field) and satisfies the homogeneous gluon field equations. So such an expansion in fermionic degrees of freedom is perfectly consistent with the idea of independent gluons, soft gluons and glueballs. However, in the self-consistent case, which we are considering here, the gluon and quark fields are interdependent and neither of these fields can be considered as independent, although it still makes sense to talk about the fermion and gluon degrees of freedom. One should also not interpret this formal expansion of the gluon field as a composite model of the gluon. It is known that the quark-antiquark interaction is repulsive in the color octet gluon channel, however, forces play no role in the operator expansion and this fact plays no role in the expansion.

By further iteration one can construct the whole operator structure of the exact quantum fields. Ultimately one arrives at an infinite sum of operator terms, each additional term containing a higher power of creation and annihilation operators. Despite the elaborate operator structure of the fields and the equations it is possible to construct a closed operator solution. In Section \ref{sec:2EQM} and Appendix \ref{sec:A}, this complete operator structure of the fields is derived and is expressed in an elegant concise form, involving a vacuum projection operator. By factoring out the quantum operators the quantum field equations can then be reduced to ordinary coupled equations, similar in structure - but not in content - to the classical equations. The elegance of this reduction is further underlined by the fact that the number of profile functions remains finite, despite the infinite nature of the expansion. The dependence on spin, color and isospin can also be factored out, leaving in the end a set of coupled differential equations for c-number profile functions. The number of profile functions depends on the symmetries present: for our QCD formulation there are only 5 gluon profile functions. But if symmetries are broken (like isospin in QED), more profile functions are needed. One of the great unexpected benefits of this quantum reduction is that the resulting non-linear equations are more amenable to analytic solutions than the original classical field equations.

After further extensive manipulations of the scalar coupled equations, two simple order $\textrm{O}(\alpha_s^0)$ solutions emerge. The first one is an uninteresting "trivial" solution, in which all effective potentials vanish. The second one features non-zero potentials that become singular at a finite radius $r_0$. This absolute confining mechanism of the quark within its self-generated bag is of great beauty and is made possible by a
unique QFT mechanism not seen before in particle theory. Since this solution is enabled by the non-linear nature of the field equations and is independent of the coupling constant, it will be denoted as a topological binding mechanism. However, this solution still features some other problems. First, in the absence of a Higgs field (we set $\phi=0$) this QCD model does not contain a scale, so there is no indication whether this model refers to a real particle. A second problem is that the expectation value of the energy operator is negative, so that this energy cannot be identified with the mass of the quark. In addition, a system with negative energy will collapse to a point as this minimizes its energy. These problems have an amazing solution. By linking QCD to general relativity (GR), we can exploit the fact that GR contains a definite scale and that gravity will act as a repulsive force if the energy of the system is negative. Hence, after the system has contracted to the Planck scale the repulsive gravitational force becomes big enough to halt the collapse. By adding vacuum energy \cite {Greben10} to the system we can compensate for the negative internal energy, and eventually obtain a net positive energy that can be identified with the mass of the system. Amazingly, this mass comes out to be slightly over 3 MeV, which is close to the phenomenological values found for the light quark masses. Hence, this solution can be identified with the light quark doublet. This solution also resolves the apparent conflict between the smallness of the quark mass and its pointlike nature. Normally, a mass of 3 MeV would correspond to a ridiculous size of 65 fm, but our result of 8.8 Planck lengths lies well within observable limits.

Now that we have explained the basic elements in our formulation of the internal bound state dynamics of QFT, it is appropriate to mention earlier approaches in QFT that tried to describe localized states. Soon after the development of the Dirac equation, this equation was used to carry out bound-state calculations, for example by deriving relativistic corrections to the non-relativistic hydrogen atom calculations. The success of these calculations already suggested that QFT can be used fruitfully in bound-state problems, however, the nature of these Dirac calculations (the proton is not treated as a quantum field) shows that such a calculation cannot really be called a genuine QFT calculation. Historically, the electron also has been subjected to bound-state calculations of quantum nature. Schweber \cite{paper1} reviews efforts to describe the electron as a bound state in the period 1940-1955. Dirac \cite{paper2} made an attempt to describe an electron as a particle with a finite, perfectly conducting, surface in 1962. However, these efforts were hampered by the fact that the electroweak and strong gauge theories had not yet been developed at the time, while one can expect that the non-linearity of these non-Abelean gauge theories is essential for the construction of bound-state solutions. Other methods have been developed to handle bound-state problems within standard QFT, for example, the Bethe-Salpeter equations (\cite{Bethe}, \cite{Gell-Mann}, \cite{Gross}, \cite{Mitra}, \cite{Mitra2}), the Blankenbecler-Sugar equations \cite{Sugar}, and the Dyson-Schwinger equations \cite{paper4}. However, none of these theories could describe the internal properties of the elementary particles as the bosonic fields were treated as external fields. More recently, lattice gauge calculations have been applied to nucleons considered as assemblies of quarks \cite{Proton_mass}. However, these calculations are also not designed to give insights in the quarks and leptons themselves.

Historically, our efforts to solve the non-linear QCD field equations self-consistently started
out as an attempt to describe the binding of three quarks in a bag,
expecting to develop an alternative to the bag models popular in the eighties \cite {Bag_model}.
Solutions were found fairly quickly \cite {paper9}, however, their physical
interpretation left much to be desired as the mass/energy scale of the light quarks (a few MeV)
corresponds to very extensive objects, which are physically unacceptable. Also,
only particle states were considered initially, as was common in the field of intermediate energy physics
 \cite {Araki} and other papers dealing with pion corrections to quark bag models \cite {Bag_model}.
It soon became clear that it is difficult to apply such an approach without anti-particles consistently, and this initial approach was replaced by a formulation in which all fields are expanded in particle and
anti-particle creation and annihilation operators. When this formulation lead to the exact operator solution it also became clear that this approach cannot be used for composite systems such as three quark bags, as the operator solutions only survive if they operate on one-body state vectors. Hence, the attention switched from describing composite systems to elementary
particles and this led to the realization that these new QFT methods can be used to give a description of the internal dynamics of elementary particles, in particular quarks.

The outline of the paper is as follows. In Section \ref{sec:2EQM} we
discuss the equations of motion and derive the form of the field
operators in terms of quark creation and annihilation operators. The
formal solution of these operator equations is presented, allowing
the reduction of the field equations to ordinary differential
equations. After defining the c-number wave function components and
c-number gluon profile functions in Section \ref{sec:profile}, we
describe the form of the ordinary second-order differential
equations, controlling the behaviour of the gluon profile functions
in Section \ref{sec:gluon_field}. These equations are highly
non-linear, and at first sight look fairly untractable. However,
after introducing a reduction process in Section
\ref{sec:technical}, we can cast these equations in an elegant
symmetrical form, which allows simple solutions to zeroth order in
$\alpha_s$. The linear equations for
$\textrm{O}(\alpha_s)$-corrections are also derived. In Section
\ref{sec:4Dirac} we discuss the (Dirac-like) field equation for the
quark and the form of the potentials. In Section \ref{sec:Solutions}
a self-consistent solution of the coupled quark-gluon equations is
presented. In
Section \ref{sec:Observe} we discuss the evaluation of expectation
values for the energy, spin and colour of the quark system. These
allow us to define sufficient boundary conditions to determine the
parameters of the solutions uniquely. In Section \ref{sec:GR} we
discuss the necessity to include coupling to GR for the trivial Higgs sector.
By incorporating vacuum energy as well, one is able to make precise estimates of the quark mass and radius. In the final section
\ref{sec:7Summary} we summarize the significance of these results
and discuss a possible steps for deriving the internal properties and masses of other particles in the standard model.

\section{The Quantization of the Equations of Motion in Quantum Chromodynamics (QCD)}
\label{sec:2EQM} As stated before we start from a basic QCD Lagrangian \cite{paper6} without electro-weak and Higgs interactions:
\begin{equation}
  \label{eq:Lagrangian}
 \mathcal{L}=\bar{\psi}(\textrm{i}\gamma _\mu
 D^\mu-M)\psi-\frac{1}{4}\textbf{F}^{\mu\nu}\bullet
 \textbf{F}_{\mu\nu},
 \end{equation}
where the covariant derivative is defined as follows:
\begin{equation}
\label{eq:Dmu} D^\mu =\partial^\mu-\frac{1}{2} \textrm{i}g_s
\boldsymbol{\lambda} \bullet\textbf{A}^{\mu}.
 \end{equation}
Here the  $\lambda_a$ are the Gell-Mann $SU(3)$-matrices. The
inproducts run over 8 components:
\begin{equation}
\label{eq:AdotB} \textbf{A} \bullet\textbf{B}=\sum _{a=1}^{8}A_aB_a.
 \end{equation}
The gluon field tensor is given by the expression:
  \begin{equation}
\label{eq:F_field}
\textbf{F}^{\mu\nu}=\partial^{\mu}\textbf{A}^{\nu}-
\partial^{\nu} \textbf{A}^{\mu}+g_s \textbf{A}^{\mu}\times
\textbf{A}^{\nu}.
 \end{equation}
 Here, the $SU(3)$ vector product is defined by:
 \begin{equation}
\label{eq:AvecB} \left( \textbf{A} \times \textbf{B}\right)_a=\sum
_{b,c=1}^{8}f_{abc} A_bB_c,
 \end{equation}
where the  $f_{abc}$ are the $SU(3)$ structure constants. The
classical equations of motion for the gluon fields are given by:
\begin{equation} \label{eq:gluon}
\partial_\mu \textbf{F}^{\nu\mu}(x)=\frac{1}{2} g_s \bar{\psi}(x)\boldsymbol{\lambda}\gamma
^\nu \psi(x) +g_s \textbf{F}^{\nu\mu} (x)\times \textbf{A}_{\mu}(x),
 \end{equation}
while the equation of motion for the quark field is:
\begin{equation} \label{eq:Dirac}
\left( \textrm{i}\gamma _\mu \partial ^\mu + \frac{1}{2}
g_s\gamma_\mu \lambda_a A_a^\mu(x) -M \right )\psi(x)=0.
\end{equation}

In the standard applications of QFT, the fermion and boson fields are expanded in the eigenstates of the free Hamiltonian, while the
interaction Hamiltonian is used to calculate the scattering diagrams in the perturbative Feynman series. Its main application lies in
scattering problems, and the fermion masses are put in by hand. In the current application of QFT we start from a more basic
level where the bare fermions have zero masses (i.e. we set $M=0$), and we try to solve the full set of field equations self-consistently for a particular (quark) state vector.
The quark is described in its own center-of-mass system (the origin of the coordinate system), and characterized by localized profile functions representing either the gluon fields or internal quark wave function.
The actual position of the system that describes the effective particle is thus irrelevant and only becomes of relevance when we enter the quark into the multi-particle environment and consider scattering
processes.

Quantization of the formulation is accomplished by expanding the fields in terms of creation and annihilation operators, the latter satisfying
certain (anti-) commutation relations. Initially the following expansion of the quark fields is used:
\begin{equation} \label{eq:quark_expansion}
\psi(x)=\sum _{\alpha}  b_\alpha \phi_\alpha(x)+\sum _{\alpha} d_\alpha^\dag \phi_\alpha^a (x),
\end{equation}
where $b_\alpha$ is the quark annihilation operator, $d^{\dag}_\alpha$ the anti-quark creation operator, and the profile functions $\phi_\alpha(x)$ and $\phi^a_\alpha(x)$ loosely
refer to the particle and anti-particle wave functions of the quark in its own center-of mass system (i.e. the internal wave functions). These wave functions are not known beforehand, and even their existence is uncertain, since it is not known whether a localized system description of the quark state $b_\alpha^\dag |0\rangle $ will emerge from the formulation. Since the field equations are solved self-consistently, only one space-time coordinate $x$ enters the considerations, so transitions between states are not associated with different times, and the overall state vector $b_\alpha^\dag |0\rangle $ stays the same. Hence, the creation and annihilation operators are also time independent and do not evolve over time. The role of these creation and annihilation operators in the current formulation should be distinguished from that in scattering problems, where they are often associated with a particular fixed functional behavior (a plane wave corresponding to a particular positive or negative hyperboloid). When scattering states undergo time evolution through interactions they change their functional behavior, and creation and annihilation operators defined in this way are mixed up, allowing even the mixing of creation and annihilation operators (Bogoliubov mixing \cite {Bogoliubov}). In our formulation the creation and annihilation operators refer to the actual physical state, represented by $b_\alpha^\dag |0\rangle $. Isolated particles do not change over time (unless they are unstable) and their state does not depend on where they are located, so $b_\alpha^\dag $ has no space-time dependence. Only, when we use an external representation of the particles and specify their momentum and/or positions with respect to other particles in scattering processes, could we contemplate time dependent creation and annihilation operators to represent the time evolution of the world state vector. However, even in that case a time independent representation of these operators seems more appropriate in QFT, as the continuous evolution of the physical system between transitions is well represented by the continuous profile functions. The evolution of one world state to another through transitions is of a stochastic nature, and should not be represented by changing creation and annihilation operators. Rather transition operators (e.g. Feynman diagrams) should relate initial and final states uniquely defined in terms of fully specified creation and annihilation operators, i.e. time-dependent operators. Clearly, in our formulation no Bogoliubov mixing occurs, and particle and anti-particle components maintain their unique meaning. This is also important for the later developments in the theory, when we make an important distinction in the algebraic properties of particle and anti-particle operators to restore the symmetry between these entities which is broken in the Lagrangian formulation. In our operator formulation the vacuum also maintains its unique identity as the state without any particles present.

We now continue the discussion of Eq.(\ref{eq:quark_expansion}). The index $\alpha$ covers the usual quantum numbers of the standard model: spin, color and charge (also called
flavor, but we only consider one generation of quarks). Since we limit ourselves to QCD, we can treat the pair of
quark states in each generation as a degenerate isospin doublet. The family/generation quantum number is supposed to be explained by the
multiplicity of the solutions and is therefore absent. Both isospin and total angular momentum are good quantum numbers in the present formulation, so that we can
limit the wave function space accordingly in the description of the quark state. Since the gluon fields carry color and spin,
color and spin transitions can take place, and an expansion in terms of a complete set of color and spin states is required. No
expansions in other quantum numbers, such as a principal quantum number, is appropriate since if a multiple of such states exist they
should follow as a solution from the given equations, rather than appear in the expansion.

If we use the expansion, Eq.(\ref{eq:quark_expansion}), in the gluon field equation Eq.(\ref{eq:gluon}), then the gluon field operators
also become expansions in quark creation and annihilation operators. In first instance this expansion looks like:
\begin{eqnarray}
\label{eq:gluon_expansion} A_a^\mu=\sum_{\alpha,\beta}b_\alpha^\dag
b_\beta A_{a,\alpha\beta}^{\mu,pp}+ \sum_{\alpha,\beta}d_\alpha
d_\beta^\dag A_{a,\alpha\beta}^{\mu,aa} \nonumber\\+
\sum_{\alpha,\beta}b_\alpha^\dag d_\beta^\dag
A_{a,\alpha\beta}^{\mu,pa}+ \sum_{\alpha,\beta}d_\alpha b_\beta
A_{a,\alpha\beta}^{\mu,ap}.
\end{eqnarray}
However, the reinsertion of this expression in the Dirac equation, Eq.(\ref{eq:Dirac}), leads to higher order terms in the expansion of
$\psi(x)$, etc., which in turn leads to higher order terms in the expansion of the gluon field. It would thus seem that we have to revert to perturbative expansions, as is done in standard QFT.
However, we will show below that the operator series can formally be summed to infinity, without increasing the number of profile functions.

To complete the quantization process we first recall the usual anti-commutation rules for the creation and annihilation operators:
\begin{equation} \label{eq:b_commute}
\left\{ b_\alpha,b_\beta^\dag \right\}=\delta_{\alpha,\beta},
\end{equation}
and
\begin{equation} \label{eq:d_commute}
\left\{ d_\alpha,d_\beta^\dag\right\}=\delta_{\alpha,\beta},
\end{equation}
while all other anti-commutators are zero. For free fields, when the fermion field is expanded in a complete set of
plane waves, one can use these anti-commutation rules (i.e. their continuous generalizations) to derive the equal-time anti-commutator
relations. For the current self-consistent treatment of the full set of coupled equations, such relations are absent. This shows that
the anti-commutation rules Eqs. (\ref{eq:b_commute}) and (\ref{eq:d_commute}) are more
fundamental than the equal-time anti-commutator relations. Historically one has often
assumed the opposite, namely that the equal-time anti-commutator relations are fundamental to QFT (e.g. by stressing the analogy with
the classical Poisson brackets), while Eqs. (\ref{eq:b_commute}) and (\ref{eq:d_commute}) were seen equivalent, but secondary. We have pointed out before that it is dangerous to rely on QFT on classical analogies, and this is a case in point.

Since the gluon field operator no longer commutes with the quark field operators, it is necessary to specify the order of the quark and gluon field operators
in accordance with the symmetries and physical requirements of the theory. The non-commutativity of boson and fermion field operators is not new to
QFT, for example Bjorken and Drell \cite {Bjorken} discuss it in the context of QED. However, in the current context the consequences are
must more farreaching. The anti-symmetric nature of the gluon field implies that we must replace Eq.(\ref{eq:F_field}) by:
\begin{equation}
\label{eq:F_field_symm}
\textbf{F}^{\mu\nu}=\partial^{\mu}\textbf{A}^{\nu}-
\partial^{\nu} \textbf{A}^{\mu}+\frac{g_s}{2}
\left\{ \textbf{A}^{\mu}\times
\textbf{A}^{\nu}-\textbf{A}^{\nu}\times \textbf{A}^{\mu}\right\}.
 \end{equation}
The quantized gluon field equations can be derived along similar lines and read:
\begin{eqnarray}
\label{eq:F_field_eq} -\partial_\mu \partial^\mu \textbf{A}^{\nu} +
\partial^\nu \left(\partial_\mu \textbf{A}^{\mu}\right)
=\frac{g_s }{2} \bar{\psi}\boldsymbol{\lambda} \gamma ^\nu \psi
\nonumber\\
\left.+g_s\left\{ \textbf{A}^{\mu} \times (\partial_\mu
\textbf{A}^{\nu}) -(\partial_\mu \textbf{A}^{\nu})\times
\textbf{A}^{\mu} \right\} \right.
\nonumber\\
+\frac{g_s}{2}\left\{ (\partial_\mu \textbf{A}^{\mu})\times
\textbf{A}^{\nu}  -\textbf{A}^{\nu} \times (\partial_\mu
\textbf{A}^{\mu})\right.
\nonumber\\
\left.-\textbf{A}^{\mu} \times (\partial^\nu \textbf{A}_{\mu})
+(\partial^\nu \textbf{A}_{\mu})\times \textbf{A}^{\mu}\right\}
\nonumber\\
-\frac{g_s^2}{4}\left\{ \textbf{A}_{\mu} \times (\textbf{A}^\nu
\times \textbf{A}^{\mu})-\textbf{A}_{\mu} \times (\textbf{A}^\mu
\times \textbf{A}^{\nu})\right.
\nonumber\\
\left.  -(\textbf{A}^\nu \times \textbf{A}^{\mu})\times
\textbf{A}_{\mu}  +(\textbf{A}^\mu \times \textbf{A}^{\nu})\times
\textbf{A}_{\mu}\right\}.
\end{eqnarray}
The quantized form of the Dirac equation is given by:
\begin{equation} \label{eq:Dirac_quantized}
\left( \textrm{i}\gamma _\mu \partial ^\mu -M \right )\psi+
\frac{1}{2} g_s\gamma_\mu \lambda_a \psi A_a^\mu  =0.
\end{equation}
The physical justification for the order of the $\psi$ and $A_a^\mu$ operator will be given in Section \ref{sec:4Dirac}.

Before we can derive the full operator expansion of the quark and gluon fields we need to refine the algebra of creation and annihilation operators. Expressions such
as $\bar{\psi}\boldsymbol{\lambda} \gamma ^\nu \psi$ display a disturbing asymmetry: the right-hand side operator $\psi$ contains quark
\emph{annihilation} operators, but antiquark \emph{creation} operators, implying that if such an operator is applied to the
vacuum, anti-particles can be created, but particles can not. However, we know that there should be a fundamental symmetry between
particles and anti-particles and that the decision to call a quark a particle or an anti-particle is largely arbitrary. So this
asymmetric outcome is unacceptable. It does not help to invert the role of particles and anti-particles, as then the problem would
be that particles can be formed from the vacuum and anti-particles can not. It seems that the symbolic mathematical language in
which we formulate QFT is unable to adequately express the natural symmetry between particles and anti-particles. The solution
to this problem is that we can maintain the same field theory language, as long as we invert the order of the anti-particle operators
in the complete string of operators between the bra and ket states. This rule can also be seen as giving meaning to the physical role
of anti-particles as being particles moving backwards in time. This elegant procedure restores the symmetry between particles and
anti-particles. The procedure also eliminates the artifact that QFT gives rise to an enormous vacuum energy, which is contradicted
by observation. This procedure will be indicated by the symbol $\mathbb{R}$, and will be called the $\mathbb{R}$-product. Hence, the
$\mathbb{R}$-product is a way to introduce ordering in expressions that are "timeless" and to enforce the property that particles and
anti-particles have opposite ordering. We should caution, however, that this $\mathbb{R}$-product only applies to quantum operators with
identical space-time variable $x$ (as is the case in our current application). Operators referring to different variables $x$ and $x'$
are not subject to this product, as their time-order can already be specified by the different times $t$ and $t'$.  To ensure the correct
application of the $\mathbb{R}$-product one could append a variable label to the operators, so that operators referring to different
space-time variables are not confused with those having common variables. Since the product does not apply to operators at different points,
it has no effect on most scattering series in QFT, which may explain why it has not been discovered before in the application of QFT.
The product is however already used implicitly in standard applications of QFT, when the normal product is inserted to remove unphysical vacuum terms.
We actually discovered this rule numerically before its theoretical origin was uncovered, as the coupled equations required unexplained minus signs, which could only be justified after the $\mathbb{R}$-product was introduced. The detailed explanation of this product is given in Appendix
\ref{sec:A}.

Using this $\mathbb{R}$-product one can derive a formal solution of the quantized equations of motion (Appendix
\ref{sec:B}). Instead of the expansions, Eqs. (\ref{eq:quark_expansion}) and (\ref{eq:gluon_expansion}), we get the exact expressions:
\begin{equation}
\label{eq:quark_expansion_full} \psi(x)=\Sigma_\infty \sum _{\alpha}
\left\{ b_\alpha \phi_\alpha(x)+d_\alpha^\dag \phi_\alpha^a
(x)\right\},
\end{equation}
and
\begin{eqnarray}
\label{eq:gluon_expansion_full}
A_a^\mu=\sum_{\alpha,\beta}b_\alpha^\dag \Sigma_\infty b_\beta
A_{a,\alpha\beta}^{\mu,pp}+ \sum_{\alpha,\beta}d_\alpha
\Sigma_\infty d_\beta^\dag A_{a,\alpha\beta}^{\mu,aa} \nonumber\\+
\sum_{\alpha,\beta}b_\alpha^\dag \Sigma_\infty d_\beta^\dag
A_{a,\alpha\beta}^{\mu,pa}+ \sum_{\alpha,\beta}d_\alpha
\Sigma_\infty b_\beta A_{a,\alpha\beta}^{\mu,ap},
\end{eqnarray}
where $\Sigma_\infty$ is the result of taking the operator
\begin{eqnarray}\label{eq:Sigma_n}
\Sigma_n=(1-N-\bar{N})\frac{2-N-\bar{N}}{2}\cdots\frac{n-N-\bar{N}}{n}\nonumber\\
\equiv
\left(\begin{array}{c}n-N-\bar{N}\\n\\
\end{array}\right),
 \end{eqnarray}
to the limit $n\rightarrow\infty$. The operators $N$ and $\bar{N}$
are given by:
\begin{equation}\label{eq:N}
N=\sum_{\epsilon}b_\epsilon^\dag b_\epsilon,
 \end{equation}
and
 \begin{equation}\label{eq:barN}
\bar{N}=-\sum_{\epsilon}d_\epsilon d_\epsilon^\dag.
 \end{equation}
The operator $\Sigma_\infty$ can be recognized as vacuum projection operator, as it only gives non-zero (unity) results if $N=\bar{N}=0$. As far as we know this operator has never before been
encountered in QFT. It will play an important role in the solution of the self-consistent field equations. Since this operator is surrounded by creation and annihilation
operators, it effectively acts like a one-body, rather than a vacuum, projection operator.

The detailed derivation of these relationships is given in Appendix
\ref{sec:B}. Since the space of internal states for the quark is
finite (see Section \ref{sec:profile}), this infinite expansion of
$\Sigma_\infty$ is terminated in practice after a finite number of
terms. Clearly, this is not true if the state is characterized by a
continuous degree of freedom, such as in the case of scattering, when we have to define $N$ as an integral over all momentum states. An
important property of this expansion is that the original number
of profile functions in the approximate expressions, Eqs.
(\ref{eq:quark_expansion}) and (\ref{eq:gluon_expansion}), does not
increase in the transition to the exact expressions
(\ref{eq:quark_expansion_full}) and (\ref{eq:gluon_expansion_full}).
This property is largely responsible for the practical feasibility and the
exactness of the current theory, and ultimately for the power of QFT.

We now want to illustrate the effective one-body nature of the projection operator. If one operates with $\psi(x)$ on a one-body
state one gets:
\begin{equation}\label{eq:one_body}
\psi(x)\left|b^\dag_\alpha\right|0\rangle =
\phi_\alpha(x)\left|0\right\rangle,
\end{equation}
however, for a many-body state one gets a zero result:
\begin{equation}\label{eq:many_body}
\psi(x)\left|b^\dag_{\alpha_1}\cdots
b^\dag_{\alpha_n}\right|0\rangle = 0,~~~~~~n>1,
\end{equation}
and similarly for anti-particle states. This shows that the only fermionic solutions of the coupled equations in the current formulation are of
one-particle nature (the term particle is used here as a generic term for particle or anti-particle).
This clearly demonstrates that we cannot apply this formulation to a composite problem like a proton system, as this involves at least
three quarks. The only way to generalize this approach to composite systems is to collapse the three quarks into a single color-singlet
point-particle. However, this would eliminate the QCD interaction and thus would not yield a proper description of nucleons. For leptons this might
be a possible model, however, this would require the introduction of the electro-weak interaction at this basic level, which we have not attempted yet.
Obviously, for protons and other hadronic systems the standard lattice
gauge calculations remain the best treatment \cite {Davies}.
Now that we have solved the quantum operator problem formally, we
can proceed towards the determination of the profile functions. We
will see that QFT has more surprises in store, and that the simplest
solution for the single quark system is of great beauty, and
realizes the localization of the bound system in an amazing way,
never seen before in physical theories.

\section{Expressions for the profile functions}
\label{sec:profile}
The next step after having removed the (creation- and annihilation-) operator structure in the
field equations is to remove the color and spin dependence, by introducing scalar wave function components and scalar gluon profile
functions. We start with the parametrization of the wave function of the 1S-quark state:
\begin{equation}
\label{eq:1S}
\phi_\alpha(\textbf{r},t)=\exp(-\textrm{i}Et)\frac{1}{r \sqrt{4\pi}}
\left( \begin{array}{c}
f(r)\\
-\textrm{i}\boldsymbol{\sigma}\bullet \hat{\textbf{r}}g(r)\\
\end{array} \right)
\chi_\alpha\eta_\alpha\xi_\alpha.
 \end{equation}
where $\chi_\alpha$ is the spin wave function, $\eta_\alpha$ is the
isospin wave function, and $\xi_\alpha$  is the colour wave
function. Here we assumed that the radial wave functions $f$ and $g$
are independent of spin, isospin and colour. This charge
independence is no longer valid if electromagnetic forces are
introduced, and $u$- and $d$-quarks become distinct. The
corresponding anti-quark states is given by:
\begin{equation}
\label{eq:1P}
\phi_\alpha^a(\textbf{r},t)=\exp(\textrm{i}Et)\frac{1}{r
\sqrt{4\pi}} \left( \begin{array}{c}
\textrm{i}\boldsymbol{\sigma}\bullet \hat{\textbf{r}}g(r)\\
f(r)\\
\end{array} \right)
\chi_\alpha\eta_\alpha\xi_\alpha.
 \end{equation}
Notice the absence of phase factors. Usually, one defines the spin
function for the anti-particle by expressions such as
$(-\textrm{i}\sigma_2 \chi_\alpha)$. Also, the colour and isospin
wave functions for the anti-particle belong to adjoint
representations of the relevant $SU(3)$ and $SU(2)$ representations.
However, the particle and anti-particle creation and annihilation
operators automatically take care of these phase factors. Hence,
Eq.(\ref{eq:1P}) does not correspond to the normal definition of the
anti-particle wave function, but exploits the simplification enabled
by the operator formalism to express the anti-particle wave
functions in the same wave function basis as the particle states.
The time dependence of these wave functions reflects the fact that
we describe stationary states. Although, the quarks belonging to the
higher generations will decay into the lighter quarks via several
decay mechanisms, we expect that these higher generations are
also characterized by a real positive eigenvalue $E$, and that the
decay width is determined in the scattering problem and not in the
current bound-state formalism.

The wave functions $f$ and $g$ are normalized according to:
\begin{equation}
\label{eq:Norm} \int_0^\infty dr[f(r)^2+g(r)^2]=1.
 \end{equation}
Using the wave functions Eq.(\ref{eq:1S}) and Eq.(\ref{eq:1P}), we
can now evaluate the source terms in Eq.(\ref{eq:F_field_eq}), and
thereby determine the form of the different gluon field components
in Eq.(\ref{eq:gluon_expansion_full}). We express the results in
terms of five profile functions:
\begin{equation}
\label{eq:A0pp} A^{0,pp}_{a,\alpha \beta}(x)
=A^{0,aa}_{a,\alpha\beta}(x)=\frac{F_0(r)}{g_sr}
\delta^{spin}_{\alpha\beta}\left(\lambda_a\right)_{\alpha\beta},
 \end{equation}
 \begin{equation}
\label{eq:Avpp} \textbf{A}^{pp}_{a,\alpha \beta}(x)
=-\textbf{A}^{0,aa}_{a,\alpha\beta}(x)=\frac{F(r)}{g_sr}
(\hat{\textbf{r}}\times
\boldsymbol{\sigma})_{\alpha\beta}\left(\lambda_a\right)_{\alpha\beta},
 \end{equation}
 \begin{equation}
\label{eq:A0pa} A^{0,pa}_{a,\alpha \beta}(x) =\textrm{i}
\frac{\tilde{F}_0(r)}{g_sr} (\hat{\textbf{r}}\bullet
\boldsymbol{\sigma})_{\alpha\beta}\left(\lambda_a\right)_{\alpha\beta}~\exp(2\textrm{i}Et),
 \end{equation}
 \begin{equation}
 \label{eq:A0ap} A^{0,ap}_{a,\alpha \beta}(x) =-\textrm{i}
\frac{\tilde{F}_0(r)}{g_sr} (\hat{\textbf{r}}\bullet
\boldsymbol{\sigma})_{\alpha\beta}\left(\lambda_a\right)_{\alpha\beta}~\exp(-2\textrm{i}Et),
 \end{equation}
 \begin{eqnarray}
\label{eq:Avpa} \textbf{A}^{pa}_{a,\alpha \beta}(x)
&=&\left\{\frac{F_1(r)}{g_sr}\boldsymbol{\sigma}_{\alpha\beta}
+\frac{F_2(r)}{g_sr} \hat{\textbf{r}}(\hat{\textbf{r}}\bullet
\boldsymbol{\sigma})_{\alpha\beta}\right\}
\nonumber\\
& \times &\left(\lambda_a\right)_{\alpha\beta}~\exp(2\textrm{i}Et),
 \end{eqnarray}
\begin{eqnarray}
\label{eq:Avap} \textbf{A}^{ap}_{a,\alpha \beta}(x)
&=&\left\{\frac{F_1(r)}{g_sr}\boldsymbol{\sigma}_{\alpha\beta}
+\frac{F_2(r)}{g_sr} \hat{\textbf{r}}(\hat{\textbf{r}}\bullet
\boldsymbol{\sigma})_{\alpha\beta}\right\}
\nonumber\\
& \times &\left(\lambda_a\right)_{\alpha\beta}~\exp(-2\textrm{i}Et).
 \end{eqnarray}
The dummy isospin Kronecker $\delta$-function has been suppressed
throughout. This simplification becomes clearly invalid as soon as we introduce
isospin breaking QED forces. In the next section we will derive equations for the five profile functions $F_0(r)$, $F(r)$,
$\tilde{F}_0(r)$, $F_1(r)$ and $F_2(r)$. The Dirac equation yields equations for the two functions $f$ and $g$, and are discussed in
Section \ref{sec:4Dirac}.

\section{Derivation of the scalar gluon field equations}
\label{sec:gluon_field}
After eliminating the common quantum operators one is left with equations in terms of
the remaining $SU(n)$ operators and functions. The $SU(3)$ operators
can easily be factored out using the relationship:
 \begin{equation}
\label{eq:24}
\sum_{b,c=1}^8f_{abc}\lambda_b\lambda_c=3\textrm{i}\lambda_a.
 \end{equation}
The spin operators satisfy numerous spin identities, which can be
used to remove their explicit occurrence as well. We are left with a
set of scalar equations of motion for the gluon profile functions.
To simplify our notation we introduce some short-hand notations for
the source functions. We define:
\begin{equation}
\label{eq:25} S_0=\frac{\alpha_s}{2}\frac{f^2+g^2}{r},
 \end{equation}
 \begin{equation}
\label{eq:26} S_1=\frac{\alpha_s}{2}\frac{2fg}{r},
 \end{equation}
 \begin{equation}
\label{eq:27} S_2=\frac{\alpha_s}{2}\frac{f^2-g^2}{r},
 \end{equation}
where $\alpha_s$ is the strong coupling constant $\alpha_s=g_s^2/4
\pi$. These source terms are related by the identity:
\begin{equation}
\label{eq:28} S_0^2-S_2^2=S_1^2.
 \end{equation}
We also introduce the auxiliary profile function $K$ , which is
going to play a central role in the dynamics:
 \begin{equation}
\label{eq:29} K=F_1+F_2.
 \end{equation}
The differential equation for $F_0$  reads:
\begin{eqnarray}
\label{eq:30} F_0''=S_0-\frac{6}{r^2}\tilde{F}_0
F_1(1-3F)-\frac{3}{r^2}(2Er+3F_0)
\nonumber\\
\times(2F_1^2+K^2) -\frac{6}{r^2}K(r\tilde{F}_0'-\tilde{F}_0)
 -\frac{3}{r^2}\tilde{F}_0(rK'+K).
 \end{eqnarray}
As expected, quadratic and cubic terms play an important role in
these equations. These terms express the non-linear nature of QCD,
and the self-coupling of gluons. The solution of the linearized
equation would lead to a Coulomb like potential.

For the magnetic particle-particle component $F$ we obtain:
\begin{eqnarray}
\label{eq:31} F''=S_1-\frac{3}{r^2}\tilde{F}_0F_1(2Er+3F_0)
\nonumber\\
+ \frac{2}{r^2}F(1-\frac{3}{2}F)(1-3F)
+\frac{3}{r^2}(\tilde{F}_0^2-F_2^2)(3F-1)
\nonumber\\
 -\frac{6}{r^2}K(rF_1'-F_1)
+\frac{3}{r^2}F_1(rK'+K) - \frac{18}{r^2}FF_1K.
 \end{eqnarray}
An earlier version of Eqs.(\ref{eq:30}) and (\ref{eq:31}), without
particle-anti-particle contributions, was presented in
Ref.\cite{paper9}. These equations did not lead to bag-like bound
states. However they already displayed the very interesting
non-linear behaviour, suggesting soliton-like solutions. It has
since become clear that the particle-anti-particle coupling is
essential for the appearance of singular confining potentials.

The profile function $\tilde{F}_0$, corresponding to the gluon field
operator $A^{0,ap}_{a,\alpha \beta}(x) $, satisfies the equation:
\begin{eqnarray}
\label{eq:32} \tilde{F}_0''=S_1+\frac{2}{r^2}\tilde{F}_0
\left\{(1-3F)^2-\frac{9}{2}K^2\right\}
\nonumber\\
 -\frac{6}{r^2}K(rF_0'-F_0)
-\frac{1}{r^2}(2ER+3F_0)(rK'+K)
\nonumber\\
+\frac{2}{r^2}F_1(1-3F)(2Er+3F_0).
 \end{eqnarray}
Finally, the two profile functions $F_1$ and $F_2$ appearing in
$\textbf{A}^{pa}_{a,\alpha \beta}(x)$, satisfy the equations:
    \begin{eqnarray}
\label{eq:33} F_1''=S_0+\frac{1}{r^2}(1-3F)(rK'-K)-\frac{F_1}{r^2}
\nonumber\\
 -\frac{F_1}{r^2}(2Er+3F_0)^2-\frac{1}{r^2}\tilde{F}_0(2ER+3F_0)(1-3F)
\nonumber\\
+\frac{1}{r^2}F_1(1-3F)^2-\frac{6}{r^2}F'K-\frac{9}{r^2}F_1K^2-\frac{9}{r^2}F_1^3,
 \end{eqnarray}
and
 \begin{eqnarray}
\label{eq:34}
\frac{F_2'}{r}-\frac{3F_2}{r^2}=S_2-S_0-\frac{F_2}{r^2}(2Er+3F_0)^2
\nonumber\\
 +F_1''-\frac{3F_1'}{r}+\frac{3F_1}{r^2}+\frac{9}{r^2}\tilde{F}_0^2K-(2Er+3F_0)
\nonumber\\
 \times\left(\frac{\tilde{F}_0'}{r}-\frac{\tilde{F}_0}{r^2}\right)+\frac{\tilde{F}_0}{r^2}(2Er+3F_0)(1-3F)
 \nonumber\\
 +\frac{3\tilde{F}_0}{r}\left(F_0'-\frac{F_0}{r}\right)+\frac{6}{r}F'F_2-\frac{15}{r^2}FF_2-\frac{9}{r^2}FF_1
 \nonumber\\
+\frac{3}{r}F(F_2'+3F_1')+\frac{9}{r^2}F^2(2K-F_1)+\frac{9}{r^2}F_1F_2^2.
 \end{eqnarray}
These 5 equations for the 5 profile functions do not look very
tractable, however, after introducing a number of new auxiliary
functions they take on a much more elegant form. We introduce the
alternative profile functions:
\begin{equation}
\label{eq:35} F_3=F_0+\frac{2}{3}Er
 \end{equation}
 and
 \begin{equation}
\label{eq:36} F_4=F-\frac{1}{3}.
 \end{equation}
Furthermore, it is convenient to introduce the following
expressions:
\begin{equation}
\label{eq:37} H=3F_3F_1-3F_4\tilde{F}_0
 \end{equation}
and
 \begin{equation}
\label{eq:38} G=3F_4^2-3F_1^2-\frac{1}{3}.
 \end{equation}
We then can write Eqs.(\ref{eq:30}-\ref{eq:33}) as follows:
\begin{eqnarray}
\label{eq:39} F_3''=S_0-\frac{6}{r^2}F_1H-\frac{9}{r^2}F_3K^2
-\frac{6}{r}K \tilde{F}_0'
\nonumber\\
-\frac{3}{r^2}\tilde{F}_0(rK'-K),
 \end{eqnarray}
 \begin{eqnarray}
\label{eq:40}
\tilde{F}_0''=S_1-\frac{6}{r^2}F_4H-\frac{9}{r^2}\tilde{F}_0K^2
-\frac{6}{r}K F_3'
\nonumber\\
-\frac{3}{r^2}F_3(rK'-K),
 \end{eqnarray}
 \begin{eqnarray}
\label{eq:41}
F_4''=S_1-\frac{3}{r^2}\tilde{F}_0H+\frac{3}{r^2}GF_4-\frac{9}{r^2}F_4K^2
\nonumber\\
-\frac{6}{r}K F_1'-\frac{3}{r^2}F_1(rK'-K),
 \end{eqnarray}
 \begin{eqnarray}
\label{eq:42}
F_1''=S_0-\frac{3}{r^2}F_3H+\frac{3}{r^2}GF_1-\frac{9}{r^2}F_1K^2
\nonumber\\
-\frac{6}{r}K F_4'-\frac{3}{r^2}F_4(rK'-K).
 \end{eqnarray}
This has not just simplified the equations enormously, but also has
brought out a strong symmetry between  $F_3$ and $\tilde{F}_0$  on
the one hand, and $F_1$  and $F_4$  on the other. The equation for
$F_2$, Eq.(\ref{eq:34}), can be added to Eq.(\ref{eq:33}), to yield
a very elegant algebraic expression for $K$:
\begin{equation}
\label{eq:43} K=r
\frac{rS_2-3F_3\tilde{F}_0'+3F_3'\tilde{F}_0-6F_4'F_1+6F_1'F_4}
{9F_3^2-9\tilde{F}_0^2+18F_1^2-18F_4^2}.
\end{equation}
We can derive another important relationship by differentiating the
numerator in Eq.(\ref{eq:43}), and inserting all the expressions for
the second derivatives. We obtain:
\begin{equation}
\label{eq:44}
S_0(\tilde{F}_0+2F_4)=S_1(F_3+2F_1)-\frac{1}{3}(rS_2)'.
 \end{equation}
If we combine this equation with the Dirac equation, we can derive a
simple relationship for the source functions, which is related to
current conservation. In the next section we continue our treatment
of the gluon profile functions exploiting the symmetry apparent in
the equations (\ref{eq:39}-\ref{eq:42}). These steps are necessary
to progress towards a physical solution, however, despite their
elegance these steps are of necessity rather technical. Hence, those
who want to immediately progress to the treatment of the quark field
equations can skip to Section \ref{sec:4Dirac}.

\section{Reduction of the scalar gluon field equations}
\label{sec:technical} We can exploit the symmetry of the equations
(\ref{eq:39}-\ref{eq:42}) to partly decouple the equations. To this
end we introduce further auxiliary functions:
 \begin{eqnarray}
\label{eq:45} X=F_3-\tilde{F}_0;~~~Y=F_3+\tilde{F}_0;
\nonumber\\
U=F_4-F_1;~~~V=F_4+F_1;~~~Z=K/r.
\end{eqnarray}
By adding and subtracting the equations (\ref{eq:39}-\ref{eq:42}) we
find:
\begin{equation}
\label{eq:46} X''=S_-+\frac{9}{r^2}U(VX-UY)-9Z^2X+6ZX'+3XZ',
\end{equation}
\begin{equation}
\label{eq:47} Y''=S_+-\frac{9}{r^2}V(VX-UY)-9Z^2Y-6ZY'-3YZ',
\end{equation}
\begin{eqnarray}
\label{eq:48} U''=-S_-+\frac{9}{2r^2}X(VX-UY)+\frac{U}{r^2}(9UV-1)
\nonumber\\
-9Z^2U+6ZU'+3UZ',
\end{eqnarray}
\begin{eqnarray}
\label{eq:49} V''=S_+-\frac{9}{2r^2}Y(VX-UY)+\frac{V}{r^2}(9UV-1)
\nonumber\\
-9Z^2V-6ZV'-3VZ',
\end{eqnarray}
where
 \begin{equation}
\label{eq:50} S_+=S_0+S_1;~~~~~S_-=S_0-S_1.
\end{equation}
We have partly decoupled the equations and can introduce further
simplifications by introducing the expression:
\begin{equation}
\label{eq:51} \hat{E}(r)=\exp \left(3\int_0^r dr'
\frac{K(r')}{r'}\right) =\exp \left(3\int_0^r dr' Z(r')\right).
\end{equation}
Because of the symmetries of the equations one can also define
$\hat{E}$ with a minus sign. We can invert this relation by
expressing $K$ in terms of $\hat{E}$:
\begin{equation}
\label{eq:KE} K(r)=\frac{r}{3}\frac{\hat{E}'(r)}{\hat{E}(r)}.
\end{equation}
Again this result is insensitive to the sign of $\hat{E}$. We now write:
\begin{eqnarray}
\label{eq:52}
X=\hat{E}\hat{X};~~~Y=\hat{E}^{-1}\hat{Y};~~~U=\hat{E}\hat{U};
~~~V=\hat{E}^{-1}\hat{V}.
 \end{eqnarray}
The new functions $\hat{X}$, $\hat{Y}$, $\hat{U}$ and $\hat{V}$
satisfy even simpler coupled equations, as the trailing $Z$ dependent terms have vanished:
 \begin{equation}
\label{eq:53}
\hat{X}''=\hat{S}_-+\frac{9}{r^2}\hat{U}\left(\hat{V}\hat{X}-\hat{U}\hat{Y}\right),
\end{equation}
\begin{equation}
\label{eq:54}
\hat{Y}''=\hat{S}_+-\frac{9}{r^2}\hat{V}\left(\hat{V}\hat{X}-\hat{U}\hat{Y}\right),
\end{equation}
\begin{equation}
\label{eq:55}
\hat{U}''=-\hat{S}_-+\frac{9}{2r^2}\hat{X}\left(\hat{V}\hat{X}-\hat{U}\hat{Y}\right)
+\frac{\hat{U}}{r^2}\left(9\hat{U}\hat{V}-1\right),
\end{equation}
\begin{equation}
\label{eq:56}
\hat{V}''=\hat{S}_+-\frac{9}{2r^2}\hat{Y}\left(\hat{V}\hat{X}-\hat{U}\hat{Y}\right)
+\frac{\hat{V}}{r^2}\left(9\hat{U}\hat{V}-1\right),
\end{equation}
where
 \begin{equation}
\label{eq:57} \hat{S}_+=S_+\hat{E};~~~\hat{S}_-=S_-\hat{E}^{-1}.
\end{equation}
We can also express Eq.(\ref{eq:43}) in terms of these new
functions, and obtain:
\begin{equation}
\label{eq:58}
\frac{1}{2}\left(\hat{X}\hat{Y}'-\hat{Y}\hat{X}'\right)+\hat{U}'\hat{V}-\hat{U}\hat{V}'=\frac{1}{3}rS_2.
\end{equation}
We can easily solve Eqs.(\ref{eq:53}-\ref{eq:56}) to $\rm
O(\alpha_s^0)$, by ignoring the source terms. By imposing the parity
requirements of the potentials (see Section \ref{sec:4Dirac}) we are
lead to the following solutions:
 \begin{equation}
\label{eq:Family1} \hat{X}=\hat{Y}=\beta
Er;~~~\hat{U}=\hat{V}=\alpha,~~~\alpha=0,\, \pm \frac{1}{3}.
\end{equation}
The solutions for $\alpha=\frac{1}{3}$ and $\alpha=-\frac{1}{3}$
appear to be physically equivalent, the former corresponding to the negative
sign of $\hat{E}$. Hence, we will only consider the set with
$\alpha=-1/3$. We will see in Section \ref{sec:Observe} that the
$\alpha=0$ solutions lead to infinite energy integrals, and must be
excluded. For $\alpha=-1/3$ the $|\beta|=2/3$ solutions are of special
interest. For $\beta=2/3$ we obtain a trivial solution where all
potentials are zero ($F_0=F=\tilde{F}_0=F_1=0;\, K =0;\,
\hat{E}=1$). For $\beta=-2/3$, $K$ - and thus $\hat{E}$ - is
singular at a finite radius $r_0$ (the bag solution). As a
consequence, all original profile functions $F_3$, $\tilde{F}_0$,
$F_4$ and $F_1$ are also singular at $r_0$. However, the reduced
functions $\hat{X}$, $\hat{Y}$, $\hat{U}$ and $\hat{V}$ remain
finite. Hence, our reduction of the equations in this section does
not just lead to a simplification of the equations, but it also
leads to a systematic elimination of the singularities from the
equations.

The solutions summarized in Eq.(\ref{eq:Family1}) are valid in the
absence of source terms. The existence of solutions for $\alpha_s\downarrow 0$
is an important consequence of the non-linear nature of the field
equations in the current bound-state approach. It is an illustration of the power of non-linear Gauge
theories in the construction of emergent complex structures. In the
standard perturbative approach to QFT, a theory with $\alpha_s\downarrow 0$ would make no sense, and
there certainly would be no binding.

Although terms higher order in $\alpha_s$ do not play an instrumental role in the binding mechanism, they will modify the detailed
outcome of the theory. In the following we show how these terms can be treated perturbatively, expressing the corrections in terms of
linear equations for the remaining profile functions. Combined with the calculation of observables in Section \ref{sec:Observe}, these linear equations
provide a powerful tool in analyzing the physical solutions in more detail. The combination of exact $\textrm{O}(\alpha_s^0)$-solutions,
with perturbative $\textrm{O}(\alpha_s)$-corrections might look somewhat heuristic, however, non-linear equations require novel
methods, and this is a very effective approach
The natural place to start the perturbative scheme is the set of
regularized equations Eqs.(\ref{eq:53}-\ref{eq:56}), which are
defined in terms of non-singular hatted profile functions. However,
the functions $\hat{X}$, $\hat{Y}$, $\hat{U}$ and $\hat{V}$ have
non-zero base values. For a perturbative scheme it is preferable
to start from zero functions. This can be done by reverting back to
functions like $F_0$, etc., but now defined in terms of the
hatted functions. Unfortunately, this requires a new set of auxiliary functions. However, the symmetry will not escape the reader, and the scheme remains very elegant and natural. The new function are:
\begin{eqnarray}
\label{eq:60} \hat{F}_3 =&
(\hat{X}+\hat{Y})/2;~~~~~~~~~\hat{\tilde{F}}_0&=(\hat{Y}-\hat{X})/2;
\nonumber\\
\hat{F}_4=&(\hat{U}+\hat{V})/2;~~~~~~~~~\hat{F}_1&=(\hat{V}-\hat{U})/2;
\nonumber\\ \hat{F}_0=&\hat{F}_3-\beta
Er;~~~~~~~~~\hat{F}&=\hat{F}_4 -\alpha;
\nonumber\\
\hat{H}=&3\hat{F}_3\hat{F}_1-3\hat{F}_4\hat{\tilde{F}}_0;~~~
\hat{G}&=3\hat{F}_4^2-3\hat{F}_1^2-\frac{1}{3}.
\end{eqnarray}
For $\beta=2/3$ and $ \alpha=-1/3$ one gets $K=0$ and $\hat{E}=1$, and the old functions are recovered (i.e. $\hat{F}_0\rightarrow F_0$.
Expressing the differential equations into the new auxiliary functions, we are led back to differential equations of
the familiar form (\ref{eq:39}-\ref{eq:42}), but now without the
$K$-contributions:
\begin{eqnarray}
\label{eq:61} \hat{F}_3''=\hat{S}_0-\frac{6}{r^2}\hat{F}_1\hat{H},
 \end{eqnarray}
 \begin{eqnarray}
\label{eq:62}
\hat{\tilde{F}}_0''=\hat{S}_1-\frac{6}{r^2}\hat{F}_4\hat{H},
 \end{eqnarray}
 \begin{eqnarray}
\label{eq:63}
\hat{F}_4''=\hat{S}_1-\frac{3}{r^2}\hat{\tilde{F}}_0\hat{H}
+\frac{3}{r^2}\hat{F}_4~\hat{G},
 \end{eqnarray}
 \begin{eqnarray}
\label{eq:64} \hat{F}_1''=\hat{S}_0-\frac{3}{r^2}\hat{F}_3\hat{H}
+\frac{3}{r^2}\hat{F}_1~\hat{G},
 \end{eqnarray}
where the reduced source functions are defined by:
\begin{equation}
\label{eq:65}
\hat{S}_0=(\hat{S}_++\hat{S}_-)/2;~~~\hat{S}_1=(\hat{S}_+-\hat{S}_-)/2.
\end{equation}
In terms of these new functions the $\rm O(\alpha_s^0)$-solutions
Eq.(\ref{eq:Family1}) are given by:
 \begin{eqnarray}
\label{eq:66} \hat{F}_3=\beta Er; ~~~\hat{F}_4=\alpha
;~~~\hat{\tilde{F}}_0= \hat{F}_1=\hat{F}_0=\hat{F}=0.
\end{eqnarray}
where $\alpha=0, \pm\frac{1}{3}$ as before. We now expand the
differential equations to first order, setting $\alpha=-1/3$. The
$\rm O(\alpha_s)$ linearized equations now read:
\begin{equation}
\label{eq:68} \hat{F}_0''=\hat{S}_0,
\end{equation}
\begin{equation}
\label{eq:69} \hat{\tilde{F}}_0''=\hat{S}_1+\frac{6\beta
E}{r}\hat{F}_1+\frac{2}{r^2}\hat{\tilde{F}}_0,
\end{equation}
\begin{equation}
\label{eq:70} \hat{F}''=\hat{S}_1+\frac{2}{r^2}\hat{F},
\end{equation}
\begin{equation}
\label{eq:71} \hat{F}_1''=\hat{S}_0-9
\beta^2E^2\hat{F}_1-\frac{3\beta E}{r}\hat{\tilde{F}}_0.
\end{equation}
These equations can be solved formally in terms of the source
functions. For the two uncoupled equations we obtain:
\begin{equation}
\label{eq:72}
\hat{F}_0=\lambda_0r-r\int_r^{r_0}dr'\hat{S}_0-\int_0^{r}dr'~r'~\hat{S}_0,
\end{equation}
and
\begin{equation}
\label{eq:73} \hat{F}=\mu
r^2-\frac{r^2}{3}\int_r^{r_0}dr'\frac{\hat{S}_1}{r'}-\frac{1}{3r}
\int_0^{r}dr'~r'^2~\hat{S}_1,
\end{equation}
where we assumed that the physical solutions have a finite spatial
extent specified by $r_0$. The parameter $\lambda_0$ could be
combined with $\beta$ into a single parameter. However, we prefer to
specify the $\rm O(\alpha_s^0)$-solution by a fixed value of
$\beta$, and treat the $\rm O(\alpha_s)$-corrections via
$\lambda_0$. The other two equations can be decoupled. First we
multiply Eq.(\ref{eq:69}) with $r$, and then differentiate the
result twice. After eliminating the function $\hat{F}_1$, we can
express the result as follows:
\begin{equation}
\label{eq:75} P''(r)+\frac{4}{r}P'(r)+9\beta^2E^2P(r)=r^{-2}Q(r),
\end{equation}
where $P(r)$ is defined by:
 \begin{equation}
\label{eq:74} P(r)=\left[\frac{\hat{\tilde{F}}_0}{r}\right]',
\end{equation}
and $Q(r)$ is defined in terms of source functions by its
derivative:
\begin{equation}
\label{eq:76} Q'(r)=(r\hat{S}_1)''+6\beta
E\hat{S}_0+9\beta^2E^2r\hat{S}_1.
\end{equation}
One can re-express this in the original source functions, and show
that:
\begin{equation}
\label{eq:Q} Q(r)=(r\hat{S}_1)'+3\beta ErS_2+\omega,
\end{equation}
where $\omega$ is the third integration constant. The solution of
Eq.(\ref{eq:75}) can be expressed in terms of the spherical Bessel
functions $j_1$ and $n_1$:
\begin{eqnarray}
\label{eq:77}P(r)= a^2\frac{j_1(ar)}{ar}\left[\nu-
\int_r^{r_0}dr'r'n_1(ar')~Q(r')\right]
 \nonumber\\-a^2
\frac{n_1(ar)}{ar}\int_0^{r}dr'r'j_1(ar')~Q(r'),
\end{eqnarray}
where $a=3\beta E$ and $\nu$ is the fourth - and final - free
constant of the $\rm O(\alpha_s)$-solutions. For the special case
$|\beta|=\frac{2}{3}$, $Q=\omega$ to $\rm O(\alpha_s)$. For small
$r$ a non-zero $\omega$ generates a singular component $\ln(r)$ in
$P(r)$ and a component $r^2 \ln(r)-r^2$ in $\hat{\tilde{F}}_0$. The
other function $\hat{F}_1$ can be deduced directly from
Eq.(\ref{eq:69}):
\begin{eqnarray}
\label{eq:F1A} 2a\hat{F}_1'=-(r\hat{S}_1)'-a^2r^2~P(r)+Q(r)=
\nonumber\\ =-a^2r^2~P(r)+arS_2(r)+\omega.
\end{eqnarray}
We can get one expression without derivatives from Eq.(\ref{eq:69}):
\begin{eqnarray}
\label{eq:80} \frac{\hat{\tilde{F}}_0}{r}+a\hat{F}_1=
-\frac{1}{2}(r\hat{S}_1)
+\frac{r^2}{2}\left[P'(r)+\frac{2}{r}P(r)\right].
\end{eqnarray}
At this stage we have four free parameters: $\lambda_0$, $\mu$,
$\nu$ and $\omega$, with dimensions $E$, $E^2$, $1$ and $E^2$,
respectively. These parameters should all be of $\rm O(\alpha_s)$.
The imposition of such a constraint on "free" parameters seems
contradictory, but the non-linear nature of the full equations
limits such freedom in the full solution. We can enforce the $\rm
O(\alpha_s)$ constraint by absorbing the free parameters into the
left-most integrals in Eqs.(\ref{eq:72}), (\ref{eq:73}) and
(\ref{eq:77}), by allowing the upper integration boundary, which is
now fixed at $r_0$, to vary between $0$ and $r_0$. In this way the
corresponding parameters are always of $\rm O(\alpha_s)$, and also
have the right order of magnitude. The need for such a prescription
is a consequence of the hybrid nature of the solution scheme: one
tries to built a perturbative solution on a non-linear $\rm
O(\alpha_s^0)$ basic solution, which itself contains a free
parameter $\beta$.

This completes the discussion of the equations for the gluon profile
functions. Although we seem to have strayed a long way from the
original gluon field tensor defined in
Eq.(\ref{eq:gluon_expansion}), expressing this tensor in terms of
the reduced profile functions is straightforward. In the next
section we discuss the Dirac field equation, which will give further
constraints on the profile and source functions. We stress again
that all profile functions and quark wave function components have
to be determined self-consistently. Hence, although we made great
strides in deriving simple equations for the gluon profile functions
in this section, we need the Dirac equation to complete the
solution.

\section{The quantized Dirac equation for isolated quarks}
\label{sec:4Dirac} We have to quantize the classical Dirac equation,
Eq.(\ref{eq:Dirac}). Since both $\psi$ and $A_a^\mu$ are operators
expressed in quark creation and annihilation operators, quantization
amounts to deciding in which order these operators have to appear in
the quantized equation. As stated in Section \ref{sec:2EQM}, the
correct choice is $\psi A_a^\mu$, leading to equation
(\ref{eq:Dirac_quantized}). This choice guarantees that the quark
interacts with the fields it generates itself. The opposite choice,
 $ A_a^\mu \psi$, does not yield a bound-state solution. For example,
 it gives rise to the
operator $b^\dag_\alpha b_\beta b_\gamma$, which vanishes when
operating on a one-particle state $
b^\dag_\delta\left|0\right\rangle$. Since two- and higher-body
terms are eliminated by the "projection" operator $\Sigma_\infty$,
none of the other terms survive either. On the other hand, $ \psi
A_a^\mu $, leads to the operator $b_\gamma b^\dag_\alpha b_\beta $,
which yields a finite result, when operating on a one-particle state
$ b^\dag_\delta\left|0\right\rangle$. Hence, the only way we can get
binding potentials and bound normalized states is to choose the
indicated order. Again we have to extract sets of terms from the
equations with similar operator form. We then can eliminate these
operators and are left with algebraic equations. In the reduction of
the operator equations we have to take into account the
$\mathbb{R}$-product when manipulating the anti-particle terms.
Ignoring this aspect would lead to erroneous signs in the reduced
equations and inconsistencies.

We can express the reduced Dirac equations as a set of scalar
equations for the large ($f$) and small ($g$) wave function
components. We find:
\begin{equation} \label{eq:82}
\left( E-V-M-V_s\right )f(r)=-\frac{g(r)}{r}-g'(r)+V_Tg(r),
\end{equation}
and
 \begin{equation} \label{eq:83}
\left( E-V+M+V_s\right )g(r)=-\frac{f(r)}{r}+f'(r)+V_Tf(r).
\end{equation}
The bare quark mass $M$ is zero in our application, however, we leave it here as it could be non-zero
when a non-zero Higgs field is introduced. Our aim is to derive the masses of dressed
quarks, so it would be inappropriate to introduce phenomenological masses at this point.

The formulation gives rise to three types of potentials (if we
introduce isospin breaking effects we would get a fourth type of
potential, as is allowed by the symmetries of the Dirac equation.
This fourth class is also of tensor type). The vector potential is
given by:
 \begin{equation} \label{eq:84}
V(r)=-\frac{C}{r}\left( F_0+2F_1\right),
\end{equation}
while the tensor potential is given by
 \begin{equation} \label{eq:85}
V_T(r)=\frac{C}{r}\left( \tilde{F}_0+2F\right).
\end{equation}
Finally, the scalar potential is exclusively given in terms of $K(r)$ :
\begin{equation} \label{eq:86}
V_s(r)=-\frac{C}{r}K(r).
\end{equation}
In these expressions $C$ is an $SU(3)$ constant with the value
$8/3$. Each of the potentials receives contributions from the
particle-anti-particle gluon profile functions. The latter terms are
essential for the emerging elegant self-consistent solutions of the
equations and thus indispensable for the existence of dressed quarks
that are finite in extent. Although the particle-anti-particle gluon
amplitudes fluctuate rapidly with time (they are proportional to
$\exp2\textrm{i}Et$ or $\exp-2\textrm{i}Et$), this time dependence
is cancelled out in the scalar Dirac equations. This is one of the
many elegant features of the current self-consistent theory. The
required behaviour of the potentials near $r=0$ fixes the
parametrization in Eq.(\ref{eq:Family1}). In lowest order $\rm
O(\alpha_s^0)$ we demand that $V$ and $V_s$ (which are even in $r$)
are constant near $r=0$, while the tensor potential (which is odd in
$r$) behaves linearly in $r$ for small $r$. This is consistent with
the small $r$-behaviour of the odd wave function $f(r)\sim r$ and
the even wave function $g(r)\sim r^2$. We will see later that in
$\rm O(\alpha_s)$ the lower component $g(r)$ also has small $r$
components $r^2~\ln(|r|)$.

By multiplying Eq.(\ref{eq:82}) with $g(r)$, and Eq.(\ref{eq:83}) by
$f(r)$, and adding them together, we can derive the following
relationship:
 \begin{equation} \label{eq:87}
\left(rS_2 \right)'=2(E-V)S_1r-2rV_TS_0+2S_0.
\end{equation}
Combining this with Eq.(\ref{eq:44}) we obtain:
 \begin{equation} \label{eq:88}
S_0~V_T+S_1~V=0,
\end{equation}
and
 \begin{equation} \label{eq:89}
2S_0~+2ErS_1=(rS_2)'.
\end{equation}
Eq.(\ref{eq:89}) also follows immediately from current conservation:
\begin{equation} \label{eq:90}
\partial_\mu \left(\bar{\psi}\gamma^\mu \psi\right)=0,
\end{equation}
if we take the particle component on the right, and the
anti-particle component on the left. Notice, that Eq.(\ref{eq:90})
is trivially satisfied for the particle-particle components. Hence,
our derivations so far have been completely consistent with current
conservation. We can derive another equation from
Eqs.(\ref{eq:82}-\ref{eq:83}) in terms of source functions:
\begin{equation} \label{eq:91}
(E-V)rS_2=rS_0(M+V_s)-\frac{1}{2}(rS_1)'.
\end{equation}
We now derive the basic non-linear self-consistent solution of the
QCD field equations by combining the conditions on the gluon and
quark profile functions.

\section{The basic self-consistent solutions for the dressed quark}
\label{sec:Solutions}
We anticipate that if a bound solution exists for the dressed quark system, that it is bound by strong,
possibly singular forces to ensure that its size is small enough to be consistent with the known properties of quarks. The
profile function $K$ may well play a prominent role in this regard, as it is given by a ratio (see Eq.(\ref{eq:43})) and thus would
become singular if the denominator turns zero. The function $K$ stands alone amongst the other profile functions in
being given by an algebraic form. This form is a unique consequence of the non-linear nature of the theory and
the inclusion of both particle and anti-particle configurations. The reduced profile functions $\hat{F}_0$, $\hat{F}$,
$\hat{\tilde{F}}_0$ and $\hat{F}_1$ satisfy regular second-order
differential equations (\ref{eq:68}-\ref{eq:71}), and are unlikely
to have singular or bag-like characteristics (we will show in the
following that the reduced source functions in these equations are
also devoid of singularities). We should note that singularities in
$K$ or $\hat{E}$ would not just be reflected in $V_s$, but also in
the other potentials, because these potentials are implicitly
defined in terms of $\hat{E}$ and $\hat{E}^{-1}$.

To analyze the behaviour of $K$ or $\hat{E}$ we cannot directly use
Eq.(\ref{eq:43}), as we have no simple equations for the unreduced
profile functions. In addition, the unreduced profile functions may
well be singular. Rather, we start by casting Eq.(\ref{eq:44}) in
reduced form:
\begin{equation}
\label{eq:122}
\frac{1}{3}(rS_2)'=-\hat{S}_0(\hat{\tilde{F}}_0+2\hat{F}_4)+\hat{S}_1(\hat{F}_3+2\hat{F}_1).
 \end{equation}
The fact that the new equation has the same structure as the
unreduced equation illustrates the elegance of the reduction
process. The advantage of the reduced form is that it is expressed
in the regularized profile functions with their known regular
$O(\alpha_s^0)$ and $O(\alpha_s)$-solutions. By re-expressing
Eq.(\ref{eq:122}) in terms of the original physical source
functions, which must be normalizable and display regular
properties, we can derive explicit expressions for $\hat{E}$ and
$\hat{E}^{-1}$:
\begin{equation}
\label{eq:hatE} \hat{E}=\frac{(rS_2)'\pm
\sqrt{\left[\left(rS_2\right)'\right]^2+S_2^2\left(A^2-B^2\right)}
}{S_+\left(A-B\right)},
\end{equation}
and
\begin{equation}
\label{eq:hatEinv} \hat{E}^{-1}=\frac{-(rS_2)'\pm
\sqrt{\left[\left(rS_2\right)'\right]^2+S_2^2\left(A^2-B^2\right)}
}{S_-\left(A+B\right)},
 \end{equation}
where
\begin{eqnarray}
\label{eq:AB} A=&3\hat{F}_3+6\hat{F}_1\rightarrow &3\beta E\,r,\nonumber\\
B=& 3\hat{\tilde{F}}_0+6\hat{F}_4\rightarrow&6\alpha \rightarrow -2.
 \end{eqnarray}
Here the right-most expressions are the $O(\alpha_s^0)$ solutions.
We can use Eq.(\ref{eq:89}) to rewrite the derivative term.

For $\beta=2/3$ and the top sign in Eqs.(\ref{eq:hatE},
\ref{eq:hatEinv}) we obtain  $\hat{E}=1$ and $K=0$, corresponding to
the trivial solution with zero potentials. The mirror solution, with
$\beta=-2/3$ and the lower (negative) sign in Eqs.(\ref{eq:hatE},
\ref{eq:hatEinv}), yields the non-trivial bag solution. For both these solutions the potential singularities
of $\hat{E}$ or $\hat{E}^{-1}$, which would be due to a zero of
$3\beta Er\pm 6\alpha$, cancel out. Notice also that the reduced
source functions remain finite, even when $\hat{E}$ or
$\hat{E}^{-1}$ become singular when $S_+$ or $S_-$ passes through
zero. Hence, the reduced gluon Eqs.(\ref{eq:68}-\ref{eq:71}) are
devoid of singularities, and form a good starting point for
calculating higher order effects.

Taking $\beta=-2/3$ we have:
\begin{equation}
\label{eq:126}
\hat{E}=\frac{S_-}{S_+}=\frac{\left(f-g\right)^2}{\left(f+g\right)^2}.
 \end{equation}
Here $\hat{E}$ is singular for $f=-g$, which serves to define the
radius of the system. This condition coincides with the usual demand
in the MIT bag model \cite{paper5} at the bag surface. Using Eq.
(\ref{eq:KE}) we obtain:
 \begin{equation}
\label{eq:127} K=\frac{4}{3}r\frac{f'g-fg'}{f^2-g^2}.
 \end{equation}
This result can also be obtained by explicit calculation from
Eq.(\ref{eq:43}). Hence, the scalar potential equals:
 \begin{equation}
\label{eq:128} V_s=-\frac{4C}{3}\frac{f'g-fg'}{f^2-g^2},
 \end{equation}
where $C = 8/3$ as before. Using Eq.(\ref{eq:126}) we can also
derive the other $O(\alpha_s^0)$-potentials:
 \begin{equation}
\label{eq:129}
V=\frac{4C}{3r}\frac{f^2+g^2}{(f^2-g^2)^2}\left[Er(f^2+g^2)+2fg\right],
 \end{equation}
and
 \begin{equation}
\label{eq:130}
V_T=-\frac{8C}{3r}\frac{fg}{(f^2-g^2)^2}\left[Er(f^2+g^2)+2fg\right].
 \end{equation}
If $r=r_0$ is the point where $f=-g$, then the quark system is
confined to $r<r_0$, as the potentials become infinite at $r_0$.
Hence, if this solution exists, then it is a bag solution with
confining potentials Eqs.(\ref{eq:128}-\ref{eq:130}).

Can we construct a solution of the Dirac equations for the
potentials Eqs.(\ref{eq:128}-\ref{eq:130})? Clearly, this is a
highly self-consistent problem, as the potentials are expressed in
the wave functions. We consider the Dirac equations
Eqs.(\ref{eq:82}-\ref{eq:83}) in more detail. First, we combine the
vector and tensor terms. Because of the identity, Eq.(\ref{eq:88}),
we can rewrite these equations as:
\begin{equation} \label{eq:131}
\left( E-M-V_s^{eff}\right )f(r)=-\frac{g(r)}{r}-g'(r),
\end{equation}
and
 \begin{equation} \label{eq:132}
\left( E+M+V_s^{eff}\right )g(r)=-\frac{f(r)}{r}+f'(r).
\end{equation}
where
 \begin{equation} \label{eq:133}
V_s^{eff}=V_s+V\frac{S_2}{S_0}=V_s-V_T\frac{S_2}{S_1}.
\end{equation}
$V_s^{eff}$ is a convenient mathematical vehicle for constructing a
solution. Notice that Eqs.(\ref{eq:131}-\ref{eq:133}) are completely
general and do not depend on the specific $O(\alpha_s^0)$ solution
considered in Eq.(\ref{eq:AB}).

Going back to the specific solution, we insert the potentials
Eqs.(\ref{eq:128}-\ref{eq:130}) into Eq.(\ref{eq:133}). We then get:
\begin{equation}
\label{eq:134}
V_s^{eff}=-\frac{4C}{3r}~\frac{r(f'g-fg')-Er(f^2+g^2)-2fg}{f^2-g^2}.
 \end{equation}
We can now eliminate $f'$ and $g'$ in this expression by means of
the effective equations Eqs.(\ref{eq:131}) and (\ref{eq:132}). For
$M=0$ one then finds $V_s^{eff}=(4C/3)\times V_s^{eff}$, which only
can be satisfied for $V_s^{eff}=0$ as $C=8/3$. This yields the free
$f$ and $g$ solutions, except that they are now confined to the
volume up to the singularity in the potentials. Re-inserting these
solutions back in $V_s^{eff}$, one finds that this effective
potential is indeed zero. Hence, we have found a self-consistent
solution of the Dirac equations for $M=0$, with the potentials
Eqs.(\ref{eq:128}-\ref{eq:130}). The solution is given by:
\begin{equation}
\label{eq:135} f(r)=N\sin(Er)
 \end{equation}
and
\begin{equation}
\label{eq:136} g(r)=N\left[\cos(Er)-\frac{\sin Er}{Er}\right].
 \end{equation}
The wave functions are defined in the range:
\begin{equation}
\label{eq:137} 0<Er<Er_0=x_0=2.04278694\cdots,
 \end{equation}
leading to a normalization constant of:
\begin{equation}
\label{eq:138} N^{-2}=\left(x_0-\frac{\sin^2x_0}{x_0}\right)E^{-1}.
 \end{equation}

\begin{figure}[!h]
\begin{center}
\includegraphics[width=8cm]{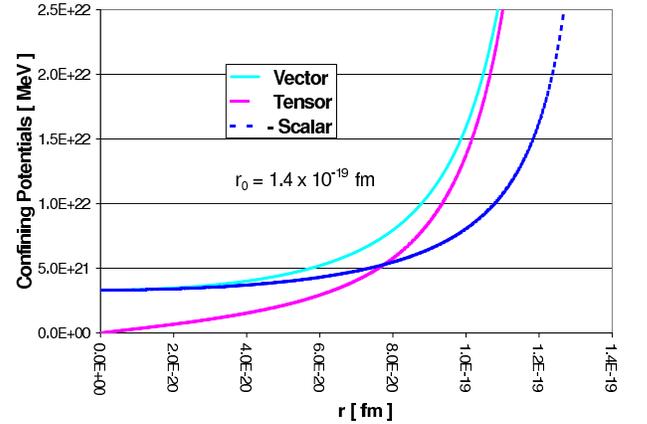}
\end{center}
\caption{Confining potentials for a $u$ or $d$ quark}
\label{fig:potentials}
\end{figure}

\begin{figure}[!h]
\begin{center}
\includegraphics[width=8cm]{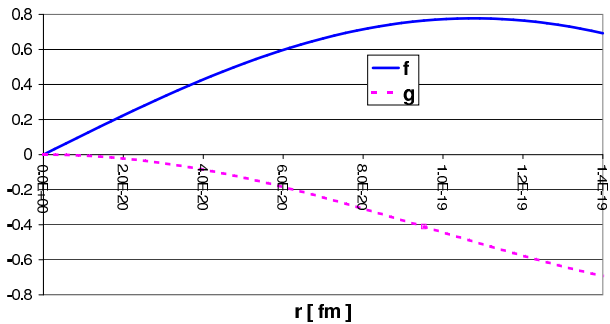}
\end{center}
\caption{Quark wave functions} \label{fig:wf}
\end{figure}

These solutions allow us to calculate the scalar, vector and tensor
potentials. In Fig. \ref{fig:potentials} we display these
potentials. The origin of the length and energy units used in this
figure will be explained in Section \ref{sec:GR}.
Clearly, the vector and tensor potentials are more singular than the
scalar one. In Fig. \ref{fig:wf} we show the resulting wave
functions, Eqs.(\ref{eq:135}-\ref{eq:138}), normalized over the
interval $[0,r_0]$.
These wave functions (in different units) also appear in the
phenomenological MIT quark bag model \cite {paper5}, where they
describe the wave functions of the three quarks in a nucleon. Hence,
the current theory shows how such a bag model could possibly
originate from fundamental QFT considerations, although the
restriction of our theory to single-particle states forbids a direct
application of our theory. Obviously, the context is also entirely
different as the size of an MIT bag is about $1$ fm, while our
single quark bag has Planck length dimensions.

We derived the solutions for a zero bare quark mass.
If we introduce a non-zero Higgs field the mass $M$ in the Dirac equation could be non-zero. In this case we find that $V_s^{eff}$ is no longer zero but becomes $V_s^{eff}=-(32/23)M$. Hence the addition
of a bare quark mass $M$ leads to a solution with an apparent mass of $M-(32/23)M=-(9/23)M$. This is an illustration of the importance
of self-consistency in these bound-state QFT solutions, and shows that any perturbative considerations may fail, since they even would
suggest the wrong sign of the mass. It shows that the calculation of quark solutions for non-zero Higgs fields is feasible, although they have important further consequences, such as additional
terms in the total energy expression.

The solutions obtained in this section also have to satisfy
additional consistency constraints. If we describe an isolated quark
by the state $\left|b^\dag_\alpha\right|0\rangle$, then its isospin,
color and spin are specified by the index $\alpha$. However, these
properties can also be defined as expectation values and expressed
as volume integrals over the quantum fields. Naturally, the two
possibilities should yield the same answer. For isospin this
condition is trivially satisfied, however, for spin and color things
are more complicated. These conditions will be discussed in Section
\ref{sec:Observe}. Our $\textrm{O}(\alpha_s^0)$ solution for
$\beta=-2/3$ satisfies the resulting conditions exactly. Hence,we
have found an exact solution of the equations of motion for
$\alpha_s=0$. From the analysis of scattering problems we know that QCD is an asymptotically free theory. Hence, it may be quite
reasonable to set $\alpha_s=0$ at the Planck scale. Since this is the scale that seems to be relevant for the light quarks (see Section \ref{sec:GR})
the $\alpha_s=0$-case might a complete description of dressed quarks in the QCD context. However, in order to describe the mass
splitting between the quarks of different charge we need to introduce QED. This theory is not asymptotically free, so that we cannot ignore this coupling
and have to apply a perturbative approach in terms of non-zero coupling constants at the Planck scale, anyway. If we assume that $\alpha_s\neq
0$, then we have to consider higher order corrections, as well and impose the boundary conditions discussed in \ref{sec:Observe}. The
color condition is the most stringent one and leads to three separate conditions in $\textrm{O}(\alpha_s)$. Hence, it effectively
fixes the integration constants $\nu$ and $\omega$, while putting indirect conditions on $\lambda_0$, which in turn leads to
$\textrm{O}(\alpha_s)$ corrections to the value of the radius $x_0$. The value of $\mu$ is essentially fixed by the spin condition.
Values for $\nu$ and $\omega$ for the bag solution are given in the next section\ref{sec:Observe}.
In that section we will also derive the formula for the total (internal) energy of a quark system. The zeroth order result will
turn out to be negative. This presents a serious problem if we want to identify this energy with the effective mass of the dressed quark.
However, there are various reasons why the model cannot be complete. In the absence of the Higgs field there is no scale parameter in QCD, hence
the scale of the quark system is undetermined and it is a mystery how the small quark mass of the light quarks can be reconciled with the effective
pointlike nature of the quarks. In Section \ref{sec:GR} we suggest a novel way to complete
the model of single quarks leading to a solution of these interconnected problems.

\section{Observables and conserved quantities}
\label{sec:Observe} The symmetries of QFT determine the conserved
quantities and observables. The invariance under phase
transformations of the quark field operator leads to current
conservation (cf. Eq.(\ref{eq:90})). This relation implies the
conservation of baryon number $B$:
\begin{equation} \label{eq:92}
B=\frac{1}{3}\int d^3x\left\langle B_\alpha
\right|\bar{\psi}\gamma_0\psi\left|B_\alpha\right\rangle,
\end{equation}
where the state vector is either a quark:
\begin{equation} \label{eq:93}
\left|B_\alpha\right\rangle=b_\alpha ^\dag\left|0\right\rangle ,
\end{equation}
or an anti-quark:
\begin{equation} \label{eq:94}
\left|B_\alpha\right\rangle=d_\alpha ^\dag\left|0\right\rangle .
\end{equation}
The normalization in Eq.(\ref{eq:92}) is chosen in such a way that
the quark has baryon number $1/3$, and the anti-quark $-1/3$. Notice
that we need the $\mathbb{R}$-product to obtain the latter
result.

The invariance under translations leads to the canonical
energy-momentum tensor:
\begin{equation} \label{eq:Tuv}
T^{\mu\nu}=\frac{\partial \mathcal{L}}{\partial \frac{\partial
\psi}{\partial x^\nu}}\frac{\partial \psi}{\partial
x_\mu}+\frac{\partial \mathcal{L}}{\partial \frac{\partial
A_a^\varrho}{\partial x^\nu}}\frac{\partial A_a^\varrho}{\partial
x_\mu}-\mathcal{L}g^{\mu\nu}.
 \end{equation}
By adding a complete differential one can re-express this as the
Gauge invariant expression (\cite {Peskin}):
\begin{equation} \label{eq:97}
\Theta^{\mu\nu}=-\mathcal{L}g^{\mu\nu}+\bar{\psi}\textrm{i}
\gamma^\nu D^\mu\psi-F_a^{\nu\rho}F_{a,\rho}^\mu,
 \end{equation}
where:
\begin{equation} \label{eq:98}
\partial_\nu T^{\mu\nu}=\partial_\nu\Theta^{\mu\nu}=0,
 \end{equation}
so that the spatial integrals of  $\Theta^{\mu 0}$ are conserved
quantities, corresponding to total energy for $\mu=0$ and total
momentum for $\mu=i$. It is easy to quantize Eq.(\ref{eq:97}) by
taking $A_{a}^\mu$ outside the quark current, symmetrizing the gluon
term, and introducing the $\mathbb{R}$-product.

The spin of the quark is related to the tensor:
 \begin{eqnarray} \label{eq:103}
\tilde{M}^{\mu\nu\rho}=M^{\mu\nu\rho}+\partial_\alpha
\left[x^\mu\left(F_a^{\rho\alpha}A_a^\nu
\right)-x^\nu\left(F_a^{\rho\alpha}A_a^\mu \right)\right]
\nonumber\\
+F_a^{\mu\rho} A_a^\nu-F_a^{\nu\rho} A_a^\mu,
\end{eqnarray}
where
\begin{eqnarray} \label{eq:102}
M^{\mu\nu\rho}=x^\mu T^{\nu\rho}-x^\nu T^{\mu\rho}+
\frac{\textrm{i}}{4}\bar{\psi}\gamma^\rho
\left(\gamma^\mu\gamma^\nu-\gamma^\nu\gamma^\mu\right)\psi
\nonumber\\
+F_a^{\mu\rho} A_a^\nu-F_a^{\nu\rho} A_a^\mu.
 \end{eqnarray}
We can show that:
\begin{equation} \label{eq:104}
\partial_\rho\tilde{M}^{\mu\nu\rho}=\partial_\rho M^{\mu\nu\rho}=0.
 \end{equation}
The spin of the quark can now be expressed by the following matrix
element:
 \begin{equation} \label{eq:109}
(S_k)_{\alpha\beta}=\frac{1}{2}\epsilon_{ijk}\int d^3r\left\langle
B_\alpha\right|\tilde{M}^{ij0}\left| B_\alpha\right\rangle.
\end{equation}
The calculated expectation value should equal $1/2$ in units
$\hbar$.

We now express the conserved quantities in terms of the gluon
profile functions and the quark wave functions or source functions.
From Eq.(\ref{eq:97}) we get for the total internal energy of the
quark system:
\begin{equation} \label{eq:105}
E_{int}=\int d^3r\left\langle B_\alpha\right|\Theta^{00}\left|
B_\alpha\right\rangle
=E-\frac{16}{9}E+\frac{16}{3\alpha_s}\int_0^{r_0} dr I(r)
 \end{equation}
where $r_0$ is the radius of the quark system and $I(r)$ is given
by:
 \begin{eqnarray} \label{eq:106}
I(r)=S_0F_3- S_1\tilde{F}_0 +\frac{1}{2}r^2
\left[\left(\frac{F_3}{r}\right)'+\frac{3}{r^2}\tilde{F}_0K\right]^2
\nonumber\\
-\frac{1}{2}r^2\left[\left(\frac{\tilde{F}_0}{r}\right)'+\frac{3}{r^2}F_3K\right]^2
- \frac{9}{r^2}\left(F_1F_3-\tilde{F}_0F_4\right)^2
\nonumber\\
 +\frac{1}{2r^2}\left(3F_1^2-3F_4^2+\frac{1}{3}\right)^2
\nonumber\\
+\frac{1}{r^2}\left(rF_4'+3F_1K\right)^2-\frac{1}{r^2}\left(rF_1'+3F_4K\right)^2.
 \end{eqnarray}
After reduction this expression simplifies:
    \begin{eqnarray} \label{eq:107}
I(r)=\hat{S}_0\hat{F}_3- \hat{S}_1\hat{\tilde{F}}_0 +\frac{1}{2}r^2
\left[\left(\frac{\hat{F}_3}{r}\right)'\right]^2
\nonumber\\
-\frac{1}{2}r^2\left[\left(\frac{\hat{\tilde{F}}_0}{r}\right)'\right]^2
-
\frac{9}{r^2}\left(\hat{F}_1\hat{F}_3-\hat{\tilde{F}}_0\hat{F}_4\right)^2
\nonumber\\
 +\frac{1}{2r^2}\left(3\hat{F}_1^2-3\hat{F}_4^2+\frac{1}{3}\right)^2
+\hat{F}_4'^2-\hat{F}_1'^2.
 \end{eqnarray}
The internal energy of the system is now expressed completely in
terms of reduced profile functions and therefore is finite even if
the original profile functions are singular. For the $\rm
O(\alpha_s^0)$ bag solution with $\beta =-2/3$ we get $E_{int} = E
-(32/9) E < 0$. In Section \ref{sec:GR} we will address the problem
associated with the negative character of this internal energy.

The total momentum of the system is easily calculated and equals
zero, as expected:
\begin{equation} \label{eq:108}
P^i_{tot}=\int d^3r\left\langle B_\alpha\right|\Theta^{i0}\left|
B_\alpha\right\rangle=0.
\end{equation}
Since we are considering a system in isolation, it would not make
sense to assign a momentum to it. Its momentum only becomes relevant
when we embed it in ordinary QFT and the quark interacts with other
particles.

Let us now consider the spin. We obtain:
 \begin{eqnarray} \label{eq:110}
(S_k)_{\alpha\beta}=(\sigma_k)_{\alpha\beta} \left\{\frac{1}{2}
-\frac{32}{9\alpha_s}\int_0^{r_0} dr
r\left(S_0F-S_1F_1\right)\right.
\nonumber\\
-\frac{32}{9\alpha_s}\int_0^{r_0} \frac{dr}{r}
\left(rF_3'-F_3+2\tilde{F}_0K\right)\left(rF_4'+2{F}_1K\right)
\nonumber\\
+\left.\frac{32}{9\alpha_s}\int_0^{r_0} \frac{dr}{r}
\left(r\tilde{F}_0'-\tilde{F}_0+2F_3K\right)\left(rF_1'+2{F}_4K\right)\right\}.
\end{eqnarray}
The first term already provides the exact answer, hence the
remaining terms should add up to zero. After reduction we can
convert this to the condition:
\begin{eqnarray} \label{eq:111}
\frac{16}{27}= -\frac{32}{9\alpha_s}\left[\int_0^{r_0} dr
r\left(\hat{S}_0\hat{F}_4-\hat{S}_1\hat{F}_1\right)\right.
\nonumber\\
-\left.\int_0^{r_0} dr
\left\{\left(r\hat{F}_3'-\hat{F}_3\right)\hat{F}_4'
-\left(r\hat{\tilde{F}}_0'-\hat{\tilde{F}}_0\right)\hat{F}_1'\right\}\right],
\end{eqnarray}
or after integration by parts:
\begin{eqnarray} \label{eq:condition1}
r_0^2 \left.\left[\left(\frac{\hat{X}}{r}\right)'\hat{V}+
\left(\frac{\hat{Y}}{r}\right)'\hat{U}\right]\right|^{r_0}=-\frac{\alpha_s}{3}.
\end{eqnarray}
To $\rm O(\alpha_s^0)$ this equation is automatically satisfied by
the solutions $\hat{X}(r)=\hat{Y}(r)=\beta Er$, e.g. by the bag
solution with $\beta =-2/3$. Its satisfaction for this bag solution
can also be related to the identity $\hat{S}_0=S_0$, which
guarantees the validity of Eq. (\ref{eq:111}) to lowest order. The
condition (\ref{eq:condition1}) can assist in fixing the four free
parameters defined in the perturbative scheme in Section
\ref{sec:technical}.

There is another conserved quantity for single quarks, namely color.
The reason that this is not usually seen as an observable, is
because quarks are usually not considered in isolation (physically
they only appear in color singlets). However, since we are
considering single quark solutions to the field equations, we
clearly have to examine this property as well. Color is associated
with the invariance of QCD under local gauge transformations. The
relevant infinitesimal transformations are \cite{paper6}:
 \begin{equation} \label{eq:113}
\psi\rightarrow\psi'=\psi-\textrm{i}\frac{\epsilon}{2}\boldsymbol{\omega}\bullet\boldsymbol{\lambda}
\psi
\end{equation}
and
 \begin{equation} \label{eq:114}
\textbf{A}_\mu\rightarrow
\textbf{A}_\mu'=\textbf{A}_\mu-\epsilon\textbf{A}_\mu\boldsymbol{\times}\boldsymbol{\omega}
-\frac{\epsilon}{g_s}\partial_\mu\boldsymbol{\omega}.
\end{equation}
The current corresponding to this transformation is given by:
 \begin{equation} \label{eq:115}
\textbf{C}^\mu=\frac{1}{2} \bar{\psi}\gamma^\mu
\boldsymbol{\lambda}\psi+\frac{1}{2}
\textbf{F}^{\mu\nu}\boldsymbol{\times}\textbf{A}_\nu-\frac{1}{2}\textbf{A}_\nu\boldsymbol{\times}\textbf{F}^{\mu\nu}.
\end{equation}
Since
 \begin{equation} \label{eq:116}
\textbf{C}^\mu=g_s^{-1}\partial_\nu \textbf{F}^{\mu\nu},
\end{equation}
it follows immediately that:
 \begin{equation} \label{eq:117}
\partial_\mu \textbf{C}^\mu=0.
\end{equation}
Hence, we can define the conserved color density ($\mu=0$):
\begin{eqnarray} \label{eq:118}
\left\langle C_k^0\right\rangle_{\alpha\beta}=\left\langle
B_\alpha\right|\int d^3x\left[\bar{\psi}\gamma^0
\frac{\lambda_k}{2}\psi\right.
\nonumber\\
+ \left.\frac{1}{2}
\left(\textbf{F}^{0\nu}\boldsymbol{\times}\textbf{A}_{\nu}
-\textbf{A}_{\nu}\boldsymbol{\times}\textbf{F}^{0\nu}\right)_k\right]\left|B_\beta\right\rangle.
\end{eqnarray}
The first term in the right-hand side of Eq.(\ref{eq:118}) equals
the bare color, so that we have to demand that the remaining terms
yield zero. Since this leads to a constraint on an integral, this
condition is not easy to implement. A simpler condition arises if we
use the identity, Eq.(\ref{eq:116}), again. This leads to the
condition:
\begin{eqnarray} \label{eq:119}
\left\langle
C_k^0\right\rangle_{\alpha\beta}=\frac{1}{g_s}\left\langle
B_\alpha\right|\int d^3x \partial_\nu
F_k^{0\nu}\left|B_\beta\right\rangle
\nonumber\\
=\frac{1}{g_s}\left\langle B_\alpha\right|\int dS
F_k^{0n}\left|B_\beta\right\rangle=
\left\langle\alpha\right|\frac{\lambda_k}{2}\left|\beta\right\rangle.
\end{eqnarray}
Writing this out in profile functions, we obtain:
 \begin{equation} \label{eq:120}
\left.r_0^2\left[\left(\frac{F_3}{r}\right)'
+\frac{3}{r^2}\tilde{F}_0K\right]\right|_{r_0}=\frac{\alpha_s}{2},
\end{equation}
where $r_0$ is the radius of the quark bag for the bag solution.
After reduction this condition can be written as:
\begin{equation} \label{eq:121}\left.
r_0^2\left[\left(\frac{\hat{X}}{r}\right)'\hat{E}
+\left(\frac{\hat{Y}}{r}\right)'\hat{E}^{-1}\right]\right|^{r_0}=\alpha_s.
\end{equation}
Notice the similarity with the spin condition Eq.
(\ref{eq:condition1}). As for the spin case, this condition is
automatically satisfied for the $\textrm{O}(\alpha_s^0)$ solutions
Eq.(\ref{eq:Family1}). We can combine the color and spin condition
to formulate the following boundary conditions:
\begin{eqnarray} \label{eq:conditions}
\left(\frac{\hat{Y}}{r}\right)'=-\frac{\alpha_s}{r_0^2}\hat{E}
\frac{1/3+\hat{E}^{-1}\hat{V}}{\hat{U}\hat{E}-\hat{E}^{-1}\hat{V}},
~~~~~~~~r\rightarrow r_0
\nonumber\\
\left(\frac{\hat{X}}{r}\right)'=\frac{\alpha_s}{r_0^2}\hat{E}^{-1}
\frac{1/3+\hat{E}\hat{U}}{\hat{U}\hat{E}-\hat{E}^{-1}\hat{V}},
~~~~~~~~r\rightarrow r_0.
\end{eqnarray}
Since either $\hat{E}$ or $\hat{E}$ is singular at $r=r_0$, we need
to treat these boundary conditions as limiting processes, rather
than as point identities. For the bag solution we have:
\begin{equation}\label{eq:144}
\hat{E}=\frac{1}{(x-x_0)^2}\tan^2 x_0
 \end{equation}
near the surface $x=x_0$. After expanding $\hat{X}$ in a Taylor
expansion for $r\rightarrow r_0$, we then arrive at three
conditions, which allow us to fix the perturbative parameters $\nu$
and $\omega$. Numerically, we find:
\begin{equation}\label{eq:nu_omega}
\nu=-0.052 \alpha_s~~~\omega=1.092 \alpha_s.
\end{equation}
The third condition is more tricky as it involves an adjustment of
the radius of the quark bag. A rough calculation indicates that we
have to adjust $x_0$ from $2.043$ to $2.129$, however, a more
complete numerical study is required to get precise answers.

In this section we considered the observables which play the role
of boundary conditions in our formulation. Other observables, such as the magnetic moment, arise from the
interactions with external fields, and must be calculated with standard perturbative methods.
Clearly, our formulation has little to add to such calculations except perhaps providing
physical cut-off parameters in view of the finite size of the dressed quarks.

\section{The role of general relativity and the determination of
the mass of the light quarks} \label{sec:GR}
As seen in the previous section, the total QFT energy of the dressed quark system is negative. Within the context
of QFT there is only one obvious contribution which could compensate for this negative energy and that is the Higgs field.
We expect that the two finite real Higgs solutions, whose classical
approximations could be characterized by the values $\phi=\pm \surd{\mu^2/\lambda}$, would have such an effect and lead to an overall positive energy,
which can be associated with the mass of the quarks. However, to enable this treatment, one needs to know the exact nature of the
Higgs Lagrangian at the fundamental level and preferably its relationship to the massive vector bosons. These Higgs terms would add an additional
set of equations and profile functions and therefore add considerably complexity. It therefore seems more natural to look initially at the trivial Higgs sector with a zero Higgs field.
Strictly speaking, this "trivial" Higgs sector is not completely zero, as the quark source term would induce a final Higgs field even if we choose the small Higgs
solution in the cubic field equation. However, we believe that it would be a fairly good approximation to approximate this lower branch by a zero Higgs field.
The consequence of this assumption is that we need to find other solutions for the following two problems (1) The absence of a scale defining parameter; (2) The negative energy of the system.
A system with negative energy and unconstrained scale would contract without bound in order
to minimize its energy, unless its scale is fixed, which it is not in QCD. To prevent this collapse we consider the consequences of
general relativity (GR). When the energy approaches the Planck scale, the effects of GR become important, and the large magnitude
of the \emph{negative} internal energy will halt the implosion. To treat these GR effects in a QFT context is non-trivial, and brings
us in uncharted territory. However, by treating these effects perturbatively, we avoid some of the tricky problems. Since this inclusion of the effects of GR,
in combination with the vacuum energy, leads to a spectacular agreement
with experiment, we feel that there is good evidence for the validity of this perturbative approach.

The effect of GR can be represented through the metric tensor, whose spatial component $\sqrt {^3g}$ modifies the spatial energy
integral:
\begin{equation}\label{eq:QFTGR}
E_{int}=\int d^3x ~\Theta_0^{~0}\rightarrow \int d^3x \sqrt
{^3g}~\Theta_0^{~0},
 \end{equation}
where the energy density is represented by the component
$\Theta_0^{~0}$ (depending on the metric convention, we may also
need an additional minus sign). The spatial metric $^3g$ is
controlled by the local energy density inside the quark. The
simplest approximation is to replace this energy distribution by an
effective mass $M_{eff}$ at the origin, with $M_{eff}$ equalling the total
internal energy $-(23/9)E$. By expanding the resulting integral in
$G$ we obtain a well-defined integral:
\begin{eqnarray}\label{eq:GRenergy}
E_{int} \approx \int d^3x
\left(1+\frac{2M_{eff}\,G}{r}\right)^{3/2}~\Theta_0^{~0}\nonumber\\
\approx \int d^3x \left(1-3\frac{\gamma GE}{r}\right)~\Theta_0^{~0},
 \end{eqnarray}
where $\gamma=23/9$. At the end of this section we will also discuss
results where the point mass at the center is replaced by an
integrated density distribution based on the internal quark wave
functions $f$ and $g$. Carrying out the integral we obtain
approximately:
\begin{equation}\label{eq:GRapprox}
E_{int}\approx - \gamma\,E \left(1- \frac{3\gamma GE^2}{ \delta
x_0}\right),
 \end{equation}
where we replaced $1/r$ by its expectation value $\left\langle 1/r
\right\rangle=E/\delta x_0$ where $\delta=0.6019\cdots$ for the bag
wave function. This expression reaches its minimum for
\begin{equation}\label{eq:scale}
E=\sqrt{\frac{\delta x_0}{9\gamma G}}~\rightarrow~r_0=\sqrt{\frac{9
x_0 \gamma G}{\delta}}.
 \end{equation}
The correction $3\gamma GE^2/\delta x_0$ to the Minkowski metric
equals $1/3$, which means that the first order (perturbative) usage
of GR is consistent. Both the radius $r_0$ and the internal
frequency $E$ are of Planck scale, with $r_0=8.8
\,\,\textrm{l}_\textrm{Planck}=1.4\times 10^{-19}\textrm{fm}$ and
$E=0.23\,\textrm{M}_\textrm{Planck}=2.8 \times 10^{21}\textrm{MeV}$.
The negativity of the internal quark energy was essential for the
stabilization of the system. Hence, a property that appeared like a
serious problem of the model actually was necessary to stabilize the
system and to fix the scale. However, the relationship between the
negative internal energy of Planck mass magnitude and the quark
mass, which is positive and lies in the MeV range, is still
unexplained.

In order to address this problem we introduce the vacuum energy
density $\epsilon$, or what is equivalent a finite cosmological
constant. Recently, a cosmological model was developed which is
based on the presence of this vacuum (dark) energy \cite{Greben10}.
The associated non-perturbative metric of the vacuum universe has a
very simple time dependence:
\begin{equation}\label{eq:guv}
g_{\mu\nu}=\eta_{\mu\nu}g(t)=\eta_{\mu\nu}(t_s/t)^2,
 \end{equation}
where the characteristic time $t_s$ follows from the postulated
vacuum energy density $\epsilon$: $t_s=(3/8\pi G\epsilon)^{1/2}$.
Another consequence of this cosmological model is that $t_s=1/H_0$,
where $H_0$ is the Hubble constant, while $t_s$ also represents the
age of the universe as measured by a co-moving observer. Fitting
recent supernovae data one arrives at a vacuum energy density of
$\epsilon=3.97\times 10^{-47}~\textrm{GeV}^4$ and an age of the
universe of $t_s=13.8$ billion years. The vacuum metric leads to the
effective time-dependent vacuum energy density:
\begin{equation}\label{eq:vacuum}
\epsilon_{eff}(t)=\sqrt{^3g(t)}~\epsilon=(t_s/t)^3~\epsilon.
 \end{equation}
We now make the novel assumption that the creation of a quark is
associated with the creation of a small vacuum universe (a small big
bang). The creation of this bag terminates at a time $t_c$, when the
size of the bag matches the size $r_0$ of the QFT quark bag. Since
this  creation time $t_c$ is very small, the effective vacuum energy
density is very large (just like it was $t_c$ seconds after the big
bang), and the resulting energy can potentially cancel the internal
energy, leaving a net quark mass of much smaller magnitude.
According to the Heisenberg uncertainty relation \cite {Briggs} the
total energy during this time interval $t_c$ is uncertain by an
amount $\hbar/2 t_c$, so that it is possible during this time
interval to create a mass of magnitude $\hbar/2 t_c$. Hence, we set:
\begin{equation}\label{eq:mq} m_q\,
=\frac{\hbar}{2}\frac{1}{t_c },
 \end{equation}
and verify whether this assumption for the quark mass is consistent
with the energy balance:
\begin{equation}\label{eq:self_consistency}
m_q=\frac{\hbar}{2}\frac{1}{t_c }=E_{vacuum}+E_{int},
 \end{equation}
where $E_{int}$ is the minimum energy determined earlier. Since the
quark mass will turn out to be infinitesimal compared to the other
terms, it is natural to ignore the left-hand term:
 \begin{equation}\label{eq:vacuum_bag}
E_{vacuum}+E_{int}=\frac{4\pi}{3}\left(\frac{x_0}{E}\right)^3\epsilon
\left(\frac{t_s}{t_c}\right)^3+E_{int}\approx 0.
 \end{equation}
Clearly, the required cancelation between internal and vacuum energy
can only happen if the internal QFT energy is negative, so the
negativity of the internal energy is again essential for the
construction of a realistic quark model. The cancelation condition
leads to a creation time $t_c$:
\begin{equation}\label{eq:tc}
t_c=\left(\frac{4\pi
x_0}{2\gamma}\right)^{1/3}\left(\frac{3}{8\pi}\right)^{1/2}\left(
\frac{G}{\epsilon}\right)^{1/6}\left(\frac{9\gamma}{\delta}\right)^{2/3}.
 \end{equation}
This expresses the creation time - and therefore the mass of the
quark - exclusively in terms of the cosmological parameters $G$ and
$\epsilon$ (or $H_0$), as the remaining parameters
($x_0=2.0478\cdots, \gamma =23/9 ~\mathrm{ and } ~\delta \approx
.60$) are fixed by the formalism. We obtain $t_c=1.04\times
10^{-22}$ s. According to Eq.(\ref{eq:mq}) this corresponds to a mass of the dressed quark of $m_q
= 3.17$ MeV, confirming the smallness of this mass in comparison to
the Planck scale. Expanding the integral in Eq.(\ref{eq:GRenergy})
to $\textrm{O}(G^2)$ only gives a small correction: one finds $m_q =
3.50$ MeV (since there are other terms contributing in
$\textrm{O}(G^2)$ which are not examined, we do not claim that this
result is more accurate than the first order result). Replacing the
mass term $M/r$ in the metric factor by a continuous distribution
based on the internal quark wave functions one finds $m_q = 4.05$
MeV. These three results give a good indication of the uncertainty
in the result $m_q = 3.17$ MeV, due to the uncertainties in unifying QCD and GR. Let us now compare this result to
"experiment".

Reference \cite {Review06} quotes the following lattice values for
the average up and down mass: $3.8 \pm 0.8$ MeV, while the average
value excluding the lattice is quoted as: $4.4 \pm 1.5$ MeV. Very
recently Davies et. al.\cite {Davies} reported on a new
determination of the light quark masses of much higher precision.
They quote the following average values for the light quark masses:
$m_q(2 ~\textrm{GeV})= 3.40 \pm .07 $ MeV and $m_q(3~ \textrm{GeV})=
3.07 \pm .06 $ MeV, where the energy in brackets is the scale used
in their calculation. The agreement between our result $3.17$ MeV
and these "experimental" values is extremely good, especially if one
takes into consideration the enormous differences between the
particle physics scale and the cosmological scale, from which our
masses have been derived. Naturally, additional effects, like the
electro-weak forces (which splits the degeneracy between u and
d-quarks) and $\textrm{O}(\alpha_s)$-terms (if $\alpha_s\neq 0$),
have to be included in more detailed calculations. However, the
current level of agreement with experiment is clearly a strong
endorsement of the theory presented here and confirms that this
solution corresponds to the first generation of quarks.

\section{Summary and Discussion}
\label{sec:7Summary}
In this paper we formulate a new set of methods in QFT that can be used to describe the internal properties and masses of elementary particles. The exact non-perturbative operator structure of the fermion and boson quantum fields is constructed by demanding the satisfaction of the full set of coupled quantum field equations. Starting from the basic QCD Lagrangian with bare pointlike massless quarks, we derive the internal wave functions of the dressed quarks and the form of the binding potentials. The binding mechanism of the localized quark-gluon state is topological in nature as it enabled by the non-linear structure of the equations and is independent of the magnitude of the strong coupling constant. Since all equations and fields refer to a single space-time coordinate, an additional operator prescription is required to reflect the opposite ordering of particles and anti-particles. This prescription (the so-called $\mathbb{R}$-product) also resolves the so-called cosmological constant problem, as it leads to a more fundamental and less ambiguous definition of the vacuum in QFT.

The calculation of the quark mass is complicated by the fact that the QFT expectation value of the energy operator is negative. It is possible that by including an elementary Higgs-quark interaction, this problem may be resolved, however, this can only account for two of the three quark generations. The third - and probably basic - generation would then correspond to the trivial (zero) Higgs solution. However, in the absence of a Higgs field we need another entity to counter the negative QFT energy and give the system a scale. We propose a novel solution for this puzzle by involving the perturbative use of gravitational forces at the Planck scale. In this way we arrive at a positive mass which is amazingly close to accepted phenomenological values for the light quarks.

The current theory would get further support if we can deduce the existence of the two higher generations of quarks. However, as we indicated above, this may well require the inclusion of a Higgs quark interaction, which will entail more profile functions and a more involved formulation. But it also offers a chance to verify various models of this force and the associated electro-weak interaction at a more fundamental level than the standard model, by comparing the predictions to experiment. Affirmation of the proposed Lagrangians would undoubtedly lead to a more unified picture of Nature. Another, desirable extension is the perturbative inclusion of QED to explain the mass differences between quarks within one generation. Clearly, QCD cannot describe the dressing of leptons, so that for this sector the non-linear electro-weak interaction must induce the binding mechanism. The application of our formulation to unbroken SU(2) is fairly straightforward (for example we already found that the factor 32/9 is replaced by 3/2), but the chiral nature of the electro-weak interaction is more difficult to include. Again, this application to leptons offers a good opportunity to test various models of the electro-weak interactions at the fundamental level, as it is possible that the complicated multiplet structure in the standard model is absent at a more fundamental level. The smallness of the neutrino masses would put extremely strong constraints on the possible theories, and if a model was successful in predicting the lepton masses then it would mark a historic breakthrough in our understanding of Nature.

The application of the new methodology is not completely limited to the description of the internal dynamics of elementary particles. It can also be used to construct free boson propagators using the new quantization method for solving the (homogeneous) equations of motion. The $\mathbb{R}$-product also could play a role in scattering diagrams if there are non-linear vertices with common space-time coordinates (as pointed out repeatedly, the $\mathbb{R}$-product does not affect the algebra of operators defined for different space-time coordinates). In scattering calculations the finite size of dressed fermions (emerging from the current theory) could also play a role in defining physical cut-offs, and thereby ensure the finiteness of otherwise infinite diagrams. However, the size of the light quarks (about 8 Planck lengths) is beyond measurement. If the higher generations of quarks are indeed due to coupling to an elementary Higgs field, then their sizes would be substantially larger, and of the order of 0.004 fm on average, so they might be measurable. Outside of these specific effects, the standard scattering formulation would remain unaltered, and so all the successful scattering applications of the standard model seem fully consistent with the current generalizations of QFT.

\appendix
\section {The $\mathbb{R}$-product}
\label{sec:A} The formulation of QFT contains an inherent asymmetry
between particles and anti-particles, which reflects the inadequacy
of the mathematical language in which we express this formulation.
Since we are constrained by this language we have to add another
rule to QFT, which together with the usual rules constitutes a
symmetric representation of the world of particles and
anti-particles. In this appendix we define this rule. It should form
an inherent part of the foundations of QFT. Its absence in the
current practice of QFT has led to many misconceptions and is also
responsible for the so-called cosmological constant problem, which
constitutes the biggest discrepancy ever between theory and
experiment in the history of physics.

Consider a typical operator expression in QFT:$\bar{\psi } O \psi$.
If we evaluate the vacuum matrix element of this operator and
concentrate on the particle contributions we get:
\begin{equation}\label{eq:A1}
<0|\bar{\psi}O \psi|0>=\sum_{\alpha,\beta}<0|b^\dag_\alpha
b_\beta|0>(\bar{\phi}_\alpha O \phi_\beta)=0,
 \end{equation}
as expected. If we look at the expectation value for a particle
state we get:
\begin{equation}\label{eq:A2}
<0|b_\gamma|\bar{\psi}O \psi|b^\dag_\gamma|0>=(\bar{\phi}_\gamma O
\phi_\gamma),
 \end{equation}
expressing the desired link between the state considered and the
matrix element. However, when we consider the anti-particle
contribution of the operator we have:
\begin{eqnarray}\label{eq:A3}
<0|\bar{\psi}O \psi|0>=\sum_{\alpha,\beta}<0|d_\alpha
d^\dag_\beta|0>(\bar{\phi}_\alpha^a O
\phi_\beta^a) \nonumber\\
=\sum_{\alpha}(\bar{\phi}_\alpha^a O \phi_\alpha^a),
 \end{eqnarray}
which is in general non-zero, and can even be infinite, as the sum
is unrestricted. Clearly, this asymmetry is unacceptable and
illustrates the incompleteness of the standard QFT description
(notice that there is no problem with mixed terms such as
$b_\alpha^\dag d_\beta^\dag$, which are responsible for the quantum
fluctuations). The origin of this asymmetry between particles and
anti-particles is that the right-hand operator $\psi$ contains
particle annihilation operators and anti-particle creation
operators. This particular combination is required by baryon (or
lepton) number conservation. However, it leads to the possibility of
creating anti-particles - and the impossibility to create particles
- when operating with $\psi$ on the ket-vector, even if it is vacuum
state. Similarly, the operator on the left, $\bar{\psi}$, contains a
combination of particle creation and anti-particle annihilation
operators, again required by baryon number conservation. This
combination displays a similar unacceptable asymmetry when operating
to the left on a bra-vector. Let us see how we can restore the
symmetry between particles and anti-particles.

To eliminate vacuum terms, like those in Eq.(\ref{eq:A3}), one
usually applies the normal product to the bilinear expression:
 \begin{equation}\label{eq:A4}
d_\alpha d_\beta^\dag \rightarrow :d_\alpha d_\beta^\dag:
=-d_\beta^\dag d_\alpha.
 \end{equation}
Clearly, the anti-particle contributions to the vacuum matrix
element now also vanish. The usual justification for this heuristic
procedure in QFT is that in QFT one is not interested in absolute
energies, but rather in relative energies, such as excitation
energies. By consistently replacing bilinear expressions by the
normal product, one then eliminates such sources of infinities from
the start. However, within GR energy leads to a modification of the
metric, so that one cannot arbitrarily rescale the energy. Thus the
heuristic replacement of bilinear expressions by their normal
product is no longer seen as acceptable. This has led to the
consideration of QFT vacuum contributions to the energy as a real
phenomenon (especially in cosmology), although this is in total
disagreement with experimental observations. As we saw above, the
normal product is needed to restore the symmetry between particles
and anti-particles, so we cannot simply dismiss it because it is a
heuristic trick. Rather, we have to explain the origin of this
heuristic prescription and should definitely not accept the reality
of the vacuum contributions as a fait accompli, especially since
these contributions are associated with an unacceptable asymmetry
between particles and anti-particles and are in strong disagreement
with observation.

The key towards the solution of this problem is to examine the
normal product from a different perspective. The normal product is
defined as a rearrangement of the operators, so that the creation
operators are shifted to the left of the annihilation operators. The
normal product is also widely used in the derivations of the
scattering expansion in QFT, although there it is used as a
convenient mathematical construct and not employed heuristically. In
its current heuristic application, Eq.(\ref{eq:A4}), it is
equivalent to the inversion of the order of the anti-particle
operators, as the normal product leaves the particle operators
unaffected in the QFT expression $\bar{\psi}O \psi$ as
$:b^\dag_\alpha b_\beta:=b^\dag_\alpha b_\beta$. Hence, an
equivalent definition of the normal product for the relevant QFT
expressions is that the order of the anti-particle operators must be
reversed. We will refer to this re-ordering as the
$\mathbb{R}$-product. We will argue that there is a fundamental
theoretical justification for this product, which is lacking in the
heuristic application of the normal product. Hence, in the bilinear
expressions considered one can justify the use of the normal product
indirectly via the justification for the $\mathbb{R}$-product. For
longer expressions the usage of the $\mathbb{R}$-product, rather
than the normal product, leads to different results. However, most
of the long expressions involve different space-time coordinates,
where neither the $\mathbb{R}$-product, nor the heuristic normal
product applies. If we analyze the usage of the heuristic
application of the normal product in the existing QFT literature, we
see that it is customary to define the (interaction) Hamiltonian
heuristically as normal ordered (see e.g. \cite {Sakurai}), before
it is inserted in the matrix elements of the $S$-matrix expansion.
This means that there is no heuristic normal ordering between
operators corresponding to different coordinates $x$ and $x'$ (it
makes no difference if we let $x\rightarrow x'$, for the current
practice it is only important that the coordinates were different
originally). In fact the heuristic usage of normal ordering between
different spatial integrals in the $S$-matrix expansion would
destroy the whole structure and meaning of the $S$-matrix expansion.
Hence, the heuristic use of the normal product must be limited to
strings of operators that all belong to a single coordinate $x$ (we
will later discuss the fundamental origin of this rule). But for
such short expressions, corresponding to a single $x$, the normal
product and the $\mathbb{R}$-product are often equivalent, so that
either one can be used. This equivalence for the simplest QFT
expressions is the likely reason why the $\mathbb{R}$-product has
not been discovered before in QFT, as the usage of the normal
product was often adequate.

For longer expressions belonging to a single space-time coordinate
the usage of the $\mathbb{R}$-product leads to very different
results than the normal product, so here it becomes of importance to
use the $\mathbb{R}$-product. In the current bound-state application
of QFT we encounter many long chains of creation and annihilation
operators belonging to a single coordinate $x$, so our transition to
the $\mathbb{R}$-product has major consequences for the current
formulation (in standard scattering applications of non-Abelian QFT
it might also have consequences, as these also involve strings of
operators at the same space-time point). Consider a longer local QFT
expression, like $d_\alpha b_\beta b_\gamma^\dag d_\delta^\dag$ and
$b_\gamma^\dag d_\delta^\dag d_\alpha b_\beta$(it is essential to
use expressions that actually occur in the usage of QFT, since we
are trying to correct a defect of QFT that occurs in its actual
representation). Using the normal product we get:
\begin{eqnarray}\label{eq:A5}
&:d_\alpha b_\beta b_\gamma^\dag d_\delta^\dag: =b_\gamma^\dag
d_\delta^\dag d_\alpha b_\beta, \nonumber\\
&:b_\gamma^\dag d_\delta^\dag d_\alpha b_\beta: =b_\gamma^\dag
d_\delta^\dag d_\alpha b_\beta,
 \end{eqnarray}
whereas the $\mathbb{R}$-product leads to:
\begin{eqnarray}\label{eq:A6}
&\mathbb{R}\left[d_\alpha b_\beta b_\gamma^\dag d_\delta^\dag\right]
=-b_\beta b_\gamma^\dag d_\delta^\dag d_\alpha,\nonumber\\
&\mathbb{R}\left[ b_\gamma^\dag d_\delta^\dag d_\alpha
b_\beta\right] =-b_\gamma^\dag b_\beta  d_\alpha d_\delta^\dag.
 \end{eqnarray}
Under the normal product the distinction between the two initial
expressions is lost, while the  $\mathbb{R}$-product maintains the
integrity of the expression, while still ensuring that the vacuum
matrix element vanishes. If we apply the resulting expressions to a
one-particle state then we get zero under the normal product, while
the $\mathbb{R}$-product gives a finite outcome. Hence, the
$\mathbb{R}$-product leads to the survival of non-linear terms in
the field equations which are essential for the existence of bound
states, such as considered in this paper.

The physical justification of the $\mathbb{R}$-product is that it
embodies Feynman's interpretation of anti-particles as particles
moving backward in time. In the $S$-matrix expansion the ordering of
operators defined at different times is taken care off by the
time-ordering procedures of standard QFT. However, when
anti-particles all belong to the same time $t$, this ordering cannot
be implemented by explicit time ordering. That is why one has to
impose the $\mathbb{R}$-product on local expressions to ensure the
correct ordering of interactions and vertices. The terms forward and
backward in time may seem a bit inappropriate for local operators
defined at a single $t$, so let us analyze this concept in the
current context. In the matrix elements the ket-vector is usually
seen as the initial state, while the bra-vector is seen as the final
state. From this perspective, the operator right-most in a chain
first operates on the ket-vector, and is "earlier" in time than
operators further to the left. For a chain of anti-particle operator
the left-most operator must first act on the initial state (i.e. the
ket vector), etc. This can be accomplished by applying the
$\mathbb{R}$-product to expressions dependent on one space-time
coordinate.

One consequence of the $\mathbb{R}$-product is that one cannot
insert a complete set of states somewhere inside the operator
expression. Hence, the manipulation and reduction of an
$\mathbb{R}$-product requires special care, as we will also see in
the following example. Consider the following reduction:
\begin{eqnarray}\label{eq:A7}
\mathbb{R}\left[b_\alpha^\dag d_\beta^\dag d_\gamma b_\delta\right]
 =-b_\alpha^\dag d_\gamma d_\beta^\dag b_\delta=-b_\alpha^\dag
 b_\delta\delta_{\gamma\beta}
 \nonumber\\
 +b_\alpha^\dag d_\beta^\dag d_\gamma b_\delta=-\delta_{\gamma\beta}
\mathbb{R}\left[b_\alpha^\dag b_\delta\right]
-\mathbb{R}\left[b_\alpha^\dag d_\gamma d_\beta^\dag
b_\delta\right].
 \end{eqnarray}
We see that if we commute the anti-particle operators under the
$\mathbb{R}$-product, we get an extra minus sign in the
anti-commutator, compared to the original anti-commutator,
Eq.(\ref{eq:d_commute}). This is conveniently summarized by the
following anti-commutation rule:
 \begin{eqnarray}\label{eq:A8}
 \mathbb{R}\left[\cdots \left\{d_\alpha^\dag, d_\beta \right\} \cdots\right]
 =\mathbb{R}\left[\cdots (-\delta_{\alpha \beta})
 \cdots\right],x=x',
 \end{eqnarray}
where $x$ and $x'$ are the coordinates belonging to the two
anti-particle operators. Since, all operators have to be expressed
in $\mathbb{R}$-product form, we have to apply this extra minus sign
whenever we contract anti-particle terms corresponding to the same
space-time point in the equations of motion or in the expressions
for the observables. These minus signs are absolutely necessary for
deriving consistent equations, in fact we established the need for
these signs, before we had discovered their origin in the
$\mathbb{R}$-product. We will see in Appendix \ref{sec:B}, that the
$\mathbb{R}$-product also enables us to derive the general operator
solution of the equations of motion by including operator chains of
any length in the expressions.

\section {General operator solution of the equations of motion}
\label{sec:B} Since we are only interested in the creation and
annihilation operator aspects in the current appendix, we will use a
simplified, partly symbolic, notation. Consider first the Dirac
equation. Writing this in the form:
\begin{equation}\label{eq:B1}
H_0\psi=\psi A,
 \end{equation}
we get a series of one- and many-body terms on the right, when we
use the expansions Eq.(\ref{eq:quark_expansion}) and
(\ref{eq:gluon_expansion}) for the quark and gluon field operators:
 \begin{eqnarray}\label{eq:B2}
H_0\psi= \sum_{\alpha,\beta,\gamma}\left[
b_\alpha\phi_\alpha+d_\alpha^\dag \phi_\alpha^a \right]\left[
b^\dag_\beta b_\gamma A_{a,\beta\gamma}^{\mu,pp}\right.
\nonumber\\
+\left. d_\beta d_\gamma^\dag
A_{a,\beta\gamma}^{\mu,aa}+b_\beta^\dag d_\gamma^\dag
A_{a,\beta\gamma}^{\mu,pa}+d_\beta b_\gamma
A_{a,\beta\gamma}^{\mu,ap}\right].
 \end{eqnarray}
After contraction, some of the right-hand terms lead back to the
single particle annihilation or anti-particle creation operator. For
example,
 \begin{equation}\label{eq:B3}
b_\alpha b_\beta^\dag
b_\gamma\rightarrow\delta_{\alpha\beta}b_\gamma;~~~~ d_\alpha^\dag
d_\beta b_\gamma\rightarrow-\delta_{\alpha\beta}b_\gamma,
 \end{equation}
where the minus sign arises from the $\mathbb{R}$-product. However,
there are also a lot of many-body operators. For example, we have:
\begin{equation}\label{eq:B4}
b_\alpha b_\beta^\dag b_\gamma\rightarrow -b_\beta^\dag b_\alpha
b_\gamma;~~~~ d_\alpha^\dag d_\beta b_\gamma\rightarrow-d_\beta
d_\alpha^\dag  b_\gamma \,,
 \end{equation}
both of which correspond to many-body terms. Notice that when
reducing operator expressions in $\mathbb{R}$-product form, the
anti-particle creation operators have to end up towards the right
after contraction. After, carrying out the $\mathbb{R}$-product in
the final reduced expression, these creation operators will then end
up to the left, as desired in reduction processes. This explains why
the second term in Eq.(\ref{eq:B4}) must be considered a many-body
term.

Writing out Eq.(\ref{eq:B2}) explicitly, we can cast the quark field
operator in the following form:
\begin{eqnarray}\label{eq:B5}
\psi= \sum_{\alpha}b_\alpha\phi_\alpha +\sum_{\alpha}d_\alpha^\dag
\phi_\alpha^a +\sum_{\alpha,\beta,\gamma}\left[b_\alpha^\dag b_\beta
b_\gamma \phi_{\alpha\beta\gamma}^{ppp}\right.
\nonumber\\
+b_\alpha d_\beta d_\gamma^\dag \phi_{\alpha\beta\gamma}^{paa}
+b_\alpha^\dag b_\beta d_\gamma^\dag \phi_{\alpha\beta\gamma}^{ppa}
+b_\alpha d_\beta b_\gamma^\dag \phi_{\alpha\beta\gamma}^{pap}
\nonumber\\
+\left. d_\alpha d_\beta^\dag d_\gamma^\dag
\phi_{\alpha\beta\gamma}^{aaa}+d_\alpha^\dag b_\beta^\dag
d_\gamma^\dag \phi_{\alpha\beta\gamma}^{apa}\right].
 \end{eqnarray}
The reason that we have $6$ many-body terms instead of the expected
$4\times 2 = 8$, is that two terms, namely $b_\alpha^\dag b_\beta
d_\gamma^\dag$ and $b_\alpha d_\beta d_\gamma^\dag$, receive
contributions from two products. In order to have unique profile
functions we demand that
\begin{equation}\label{eq:B6}
\phi_{\alpha\beta\gamma}^{ppp}=-\phi_{\alpha\gamma\beta}^{ppp};~~
\phi_{\alpha\beta\gamma}^{aaa}=-\phi_{\alpha\gamma\beta}^{aaa}~~
\mathrm{and}~~
\phi_{\alpha\beta\gamma}^{apa}=-\phi_{\gamma\beta\alpha}^{apa}.
 \end{equation}
For the gluon field operator we can carry out a similar expansion,
by demanding that it has the same form as the many-body source
function. For the quartic component of the gluon field operator we
now get 16 terms, minus the duplicate products, i.e. 12 in total.
Again, we have to impose anti-symmetrization conditions on the gluon
profile functions, to ensure their uniqueness.

To simplify matters we will concentrate on the particle terms,
suppressing the particle indices for the moment. We then have:
 \begin{equation}\label{eq:B7}
A=\sum_{\alpha\beta}A_{\alpha\beta}b_\alpha^\dag b_\beta
+\sum_{\alpha\beta\gamma\delta}A_{\alpha\beta\gamma\delta}b_\alpha^\dag
b_\beta^\dag b_\gamma b_\delta,
 \end{equation}
where we demand that
\begin{equation}\label{eq:B8}
A_{\alpha\beta\gamma\delta}=-A_{\beta\alpha\gamma\delta}=
-A_{\alpha\beta\delta\gamma}=A_{\beta\alpha\delta\gamma}.
 \end{equation}
The Dirac equation for the $b_\alpha b_\beta b_\gamma$ component
then reads:
\begin{eqnarray}
\label{eq:B9}
H_0\phi_{\alpha\beta\gamma}=-\frac{1}{2}\phi_{\beta}A_{\alpha\gamma}
+\frac{1}{2}\phi_{\gamma}A_{\alpha\beta}
+2\phi_{\epsilon}A_{\epsilon\alpha\beta\gamma}
\nonumber\\
+\phi_{\alpha\beta\epsilon}A_{\epsilon\gamma}
-\phi_{\alpha\gamma\epsilon} A_{\epsilon\gamma}
+2\phi_{\alpha\epsilon\tau}A_{\tau\epsilon\beta\gamma}.
 \end{eqnarray}
Hence, even if we limit ourselves to particle terms, we still get
fairly complicated equations. Because of the first two terms on the
right we do not have a solution  $\phi_{\alpha\beta\gamma}= 0$ and
$A_{\alpha\beta\gamma\delta}= 0$. Hence, the solution of the field
equations necessarily contains many-body components. Iterating the
equations further, we can use the same argument in the next order,
and so on, ad infinitum. Hence, the solutions of the field equations
are operators that contain components that are of arbitrary high
order in the creation and annihilation operators. What is more,
these components cannot be treated perturbatively, as they are the
same order as the one-body terms. This series is
cut off in practice by the finite size of the wave function space, as products
containing two identical creation or
annihilation operators gives zero. For scattering problems the wave function space is infinite,
however, here one often deals with operators at different space-time points between which the $\mathbb{R}$-product does not feature. 

After analyzing Eq.(\ref{eq:B9}) and the corresponding quartic gluon
operator equation, we find the following solution:
\begin{equation}\label{eq:B10}
\phi_{\alpha\beta\gamma}=\frac{1}{2}\left(\phi_\beta
\delta_{\alpha\gamma}-\phi_\gamma \delta_{\alpha\beta}\right),
 \end{equation}
and
 \begin{equation}\label{eq:B11}
A_{\alpha\beta\gamma\delta}=-\frac{1}{4}\left(\delta_{\beta\gamma}A_{\alpha\delta}
+\delta_{\delta\alpha}A_{\beta\gamma}-\delta_{\alpha\gamma}A_{\beta\delta}
-\delta_{\beta\delta}A_{\alpha\gamma}\right).
 \end{equation}
We can use Eq.(\ref{eq:B10}) to write the quark field operator as
follows:
\begin{eqnarray}\label{eq:B12}
\psi=\sum_{\alpha}b_\alpha\phi_\alpha +\sum_{\alpha\beta\gamma}
b_\alpha^\dag b_\beta b_\gamma \phi_\beta\delta_{\alpha\gamma}
\nonumber\\
=(1-\sum_{\epsilon}b_\epsilon^\dag
b_\epsilon)\sum_{\alpha}b_\alpha\phi_\alpha=(1-N)\sum_{\alpha}b_\alpha\phi_\alpha,
 \end{eqnarray}
where
 \begin{equation}\label{eq:B13}
N=\sum_{\epsilon}b_\epsilon^\dag b_\epsilon,
 \end{equation}
is the particle number operator. Similarly, we can write:
 \begin{equation}\label{eq:B14}
A=\sum_{\alpha\beta}b_\alpha^\dag (1-N)b_\beta A_{\alpha\beta}.
 \end{equation}
We can show that the non-linear terms in the gluon field equations
do not invalidate this result, by noticing that:
 \begin{eqnarray}\label{eq:B15}
AA\rightarrow b_\alpha^\dag (1-N)b_\beta A_{\alpha\beta}
b_\gamma^\dag (1-N)b_\delta A_{\gamma\delta}\rightarrow \nonumber\\
b_\alpha^\dag (1-N)b_\delta A_{\alpha\beta}A_{\beta \delta}+\cdots
 \end{eqnarray}
where the $\cdots$ stand for higher-order terms, which can be
ignored at this stage. So products of $A$'s also lead to terms of
the form, Eq.(\ref{eq:B14}). Their functional form is the same as
the one-body terms considered, which ensures that the scalar field
equations have a simple connected structure without need for
perturbative solutions at the operator level.

We can extend this analysis to anti-particles. This analysis is a
bit trickier because of the implied $\mathbb{R}$-product. We find:
\begin{equation}\label{eq:B16}
\psi=(1-N-\bar{N})\sum_{\alpha}\left(b_\alpha
\phi_{\alpha}+d_\alpha^\dag\phi_\alpha^a\right),
 \end{equation}
and
\begin{eqnarray}\label{eq:B17}
A=\sum_{\alpha\beta}\left[b_\alpha^\dag(1-N-\bar{N})b_\beta
A_{\alpha\beta}^{pp}+d_\alpha(1-N-\bar{N})b_\beta
A_{\alpha\beta}^{ap}\right.
\nonumber\\
\left.+b_\alpha^\dag(1-N-\bar{N})d_\beta^\dag
A_{\alpha\beta}^{pa}+d_\alpha(1-N-\bar{N})d_\beta^\dag
A_{\alpha\beta}^{aa}\right],
\nonumber\\
 \end{eqnarray}
where
\begin{equation}\label{eq:B18}
\bar{N}=-\sum_{\epsilon}d_\epsilon d_\epsilon^\dag,
 \end{equation}
is the effective anti-particle number operator under the
$\mathbb{R}$-product. The simplicity of this result is only true
under the $\mathbb{R}$-product. Once, we carry out this product
explicitly, the elegance and simplicity is lost.

We can now consider terms of higher order, some of which already
appeared in our expressions, e.g. in Eq.(\ref{eq:B15}).
Surprisingly, we can get an exact solution of the many-body
equations to any order. We can write:
\begin{equation}\label{eq:B19}
\psi=\Sigma_\infty\sum_{\alpha}\left(b_\alpha
\phi_{\alpha}+d_\alpha^\dag\phi_\alpha^a\right).
 \end{equation}
Here we defined:
 \begin{eqnarray}\label{eq:B20}
\Sigma_n=(1-N-\bar{N})\frac{2-N-\bar{N}}{2}\cdots\frac{n-N-\bar{N}}{n}\nonumber\\
\equiv
\left(\begin{array}{c}n-N-\bar{N}\\n\\
\end{array}\right),
 \end{eqnarray}
where we generalized the definition of the factorial to allow for
operators. The gluon field operator solution becomes:
\begin{eqnarray}\label{eq:B21}
A^\mu_a=\sum_{\alpha\beta}\left[b_\alpha^\dag\Sigma_\infty b_\beta
A_{a,\alpha\beta}^{\mu,pp}+d_\alpha\Sigma_\infty b_\beta
A_{a,\alpha\beta}^{\mu,ap}\right.
\nonumber\\
\left.+b_\alpha^\dag \Sigma_\infty d_\beta^\dag
A_{a,\alpha\beta}^{\mu,pa}+d_\alpha \Sigma_\infty d_\beta^\dag
A_{a,\alpha\beta}^{\mu,aa}\right],
 \end{eqnarray}
where we re-introduced the upper and lower indices in $A$. The
amazing thing is that we have found an exact solution without
increasing the number of profile functions. These general solutions
of the operator problem allow us to reduce the operator equations to
scalar equations. The operator $\Sigma_\infty$ is a vacuum projection
operator, as it yields zero unless $N=\bar{N}=0$. However, within
operator expressions it is surrounded by single creation and annihilation operators,
which implies that in practice it acts like a one-body
projection operator. Hence, we have proved that this exact solution
of the operator field equations only yields solutions when operating on one-body state vectors. Furthermore,
the quark creation operator operating on the vacuum, which
originally created a bare quark, now creates a dressed quark state,
and not just a component thereof. It would seem natural to represent
many-body components (like sea quarks) in the state vector. However,
as this study has shown, these many-body components appear in the
field operators, and conspire to leave the state vector simple, so
that a (dressed) quark still can be represented by a single creation
operator and the simple state vector $b_\alpha^\dag|0\rangle$.

Another consequence of this result is that these self-consistent
solutions of the field equations cannot describe multi-quark states,
such as the proton or the pion. This is a bit disappointing, since
the emergence of the MIT-bag mechanism from QFT, would suggest that
a similar derivation can be used to justify existing
phenomenological approaches to the multi-quark proton case. However,
within the present operator solution this is only feasible if the
multi-quark state is treated as a one-body state (as the proton was
before quarks were discovered). But an "elementary" proton is a
color singlet, and thus it would no longer be possible to treat it
via the non-Abelian $SU(3)$-theory, as was essential for the
emergence of the bag solution. So the existing techniques,
such as lattice gauge calculations, remain the tool of choice to
deal with multi-quark systems.

%\end{ack}
%\end{appendix}

%\begin{acknowledgments}
% put your acknowledgments here.
%The authors acknowledge the support provided by
%\end{acknowledgments}

\end{document}